\documentclass[a4paper]{ar-1col}

\usepackage{amssymb,amsfonts,amsmath,amstext,amsgen,amsopn,amsxtra,indentfirst,rotating,natbib,graphicx,subfigure}

\jname{Annual Review of Earth \& Planetary Sciences}
\jvol{43}
\jyear{2015}
\doi{10.1146/annurev-earth-060614-105146}

\begin{document}

\markboth{Heng \& Showman}{Atmospheric Dynamics of Hot Exoplanets}

\title{Atmospheric Dynamics of Hot Exoplanets}

\author{Kevin Heng$^1$ and Adam P. Showman$^2$
\affil{$^1$University of Bern, Physics Institute, Center for Space and Habitability, Sidlerstrasse 5, CH-3012, Bern, Switzerland.  Email: kevin.heng@csh.unibe.ch}
\affil{$^2$University of Arizona, Department of Planetary Sciences \& the Lunar and Planetary Laboratory, 1629 University Blvd., Tucson, AZ 85721, U.S.A.  Email: showman@lpl.arizona.edu}}

\begin{abstract}
The characterization of exoplanetary atmospheres has come of age in the last decade, as astronomical techniques now allow for albedos, chemical abundances, temperature profiles and maps, rotation periods and even wind speeds to be measured.  Atmospheric dynamics sets the background state of density, temperature and velocity that determines or influences the spectral and temporal appearance of an exoplanetary atmosphere.  Hot exoplanets are most amenable to these characterization techniques; in the present review, we focus on highly-irradiated, large exoplanets (the ``hot Jupiters"), as astronomical data begin to confront theoretical questions.  We summarize the basic atmospheric quantities inferred from the astronomical observations.  We review the state of the art by addressing a series of current questions and look towards the future by considering a separate set of exploratory questions.  Attaining the next level of understanding will require a concerted effort of constructing multi-faceted, multi-wavelength datasets for benchmark objects.  Understanding clouds presents a formidable obstacle, as they introduce degeneracies into the interpretation of spectra, yet their properties and existence are directly influenced by atmospheric dynamics.  Confronting general circulation models with these multi-faceted, multi-wavelength datasets will help us understand these and other degeneracies.
\end{abstract}

\maketitle

\tableofcontents

\section{Introduction}

\subsection{Hot and/or large exoplanets are easiest to characterize}

For millenia, our understanding of worlds beyond Earth was confined to the Solar System.  This Solar System- and Earth-centric perspective was shattered in the mid-1990s when astronomers began finding planets orbiting other stars---first around pulsars \citep{wf92}, then later around a Sun-like star \citep{mq95}.  With an ever-expanding database of thousands of exoplanets and exoplanet candidates, the discovery of planets beyond our Solar System is a firmly established enterprise; exoplanet characterization is a nascent, flourishing field.  Understanding exoplanets and their atmospheres may be the only path towards investigating the possibility of life elsewhere in the Universe.  Among the several techniques available for exoplanet detection, the radial velocity and transit methods stand out as the work horses, having been used to detect the majority of these exoplanets.  The former measures the wobble of the star as it and its exoplanets orbit around their common center of mass.  The latter measures the diminution of light as exoplanets pass in front of (the ``transit") and behind (the ``secondary eclipse"\footnote{Some workers use the terms ``eclipse" or ``occultation".}) the star.  Used in tandem, they yield the mass and radius of an exoplanet.  Both methods require the star to be fairly quiescent, as stellar flares and star spots are sources of contamination and confusion.

\begin{figure}
\begin{center}
\includegraphics[width=0.75\columnwidth]{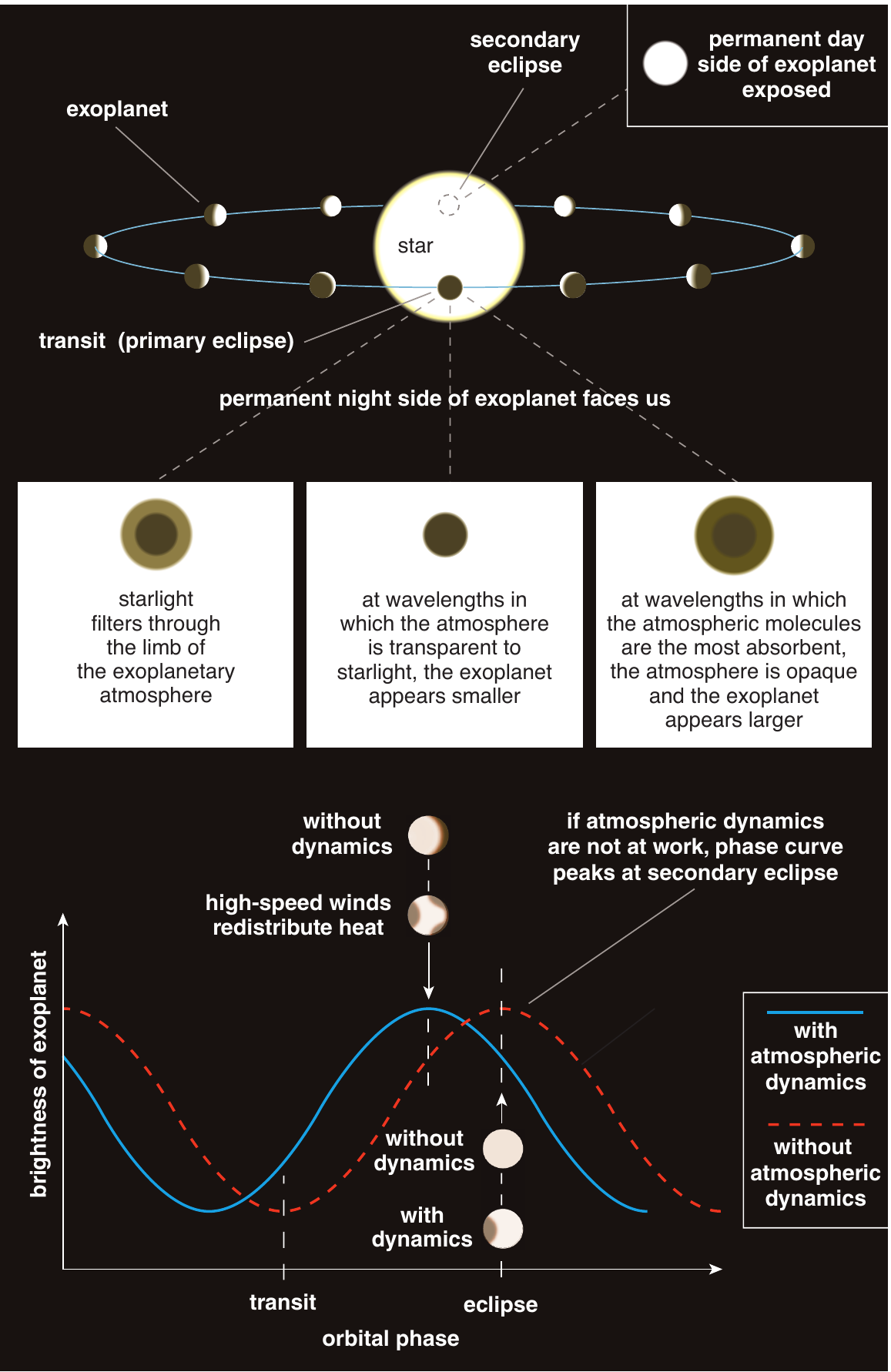}
\end{center}
\caption{Obtaining atmospheric information from transiting exoplanets via transits, eclipses and phase curves.  Courtesy of Tom Dunne (taken from \citealt{heng12b}).}
\label{fig:schematic0}
\end{figure}

Transits measured at multiple wavelengths also provide for a powerful way of characterizing the atmospheres of exoplanets (Figure \ref{fig:schematic0}).  Atoms and molecules present in the atmosphere absorb or scatter starlight to different degrees across wavelength, which translates into a variation in the radius of an exoplanet with wavelength.  By measuring the wavelength-dependent radius, one obtains a transmission spectrum from which atmospheric composition may be inferred.  However, to measure a transit in the first place requires that the exoplanet resides in a nearly edge-on orbit.  Geometry dictates that this is more likely to occur for exoplanets residing closer ($\sim 0.1$ AU)\footnote{The Earth-Sun distance is the ``astronomical unit", denoted by ``AU".} to their stars, since the transit probability is roughly the ratio of the stellar radius to the exoplanet-star separation.  The variation in radius is also easier to measure for larger exoplanets, since the method actually measures the ratio of the exoplanetary to the stellar radius.  Hot, close-in exoplanets also radiate higher heat fluxes than their cooler counterparts, implying larger secondary eclipse depths and phase curve variations in the infrared.  For these reasons, hot, large\footnote{Once the transits of an exoplanet are detected, the quantification of spectral features is easier for exoplanets with lower densities \citep{dewit13}.} exoplanets are the most amenable to atmospheric characterization via the transit method.  One of the surprises of the exoplanet hunt was the discovery of ``hot Jupiters": Jupiter-like exoplanets located $\sim 0.1$ AU from their stars with atmospheric temperatures $\sim 1000$--3000 K \citep{mq95}.  Hot Jupiters provide unprecedented laboratories for the characterization of exoplanetary atmospheres.

An emerging and complementary method is direct imaging, which---as the name suggests---obtains images of the exoplanet by isolating them from the accompanying starlight via coronographic methods and/or novel image reduction techniques.  The most iconic examples are the four directly imaged exoplanets in the HR 8799 system \citep{marois08,marois10}.  As one may expect, this technique works best when the exoplanets are hot (and therefore brighter at shorter wavelengths, as dictated by Wien's law) and located far away ($\sim 10$--100 AU) from their stars.  Directly imaged exoplanets are hot not because of stellar irradiation (which is a negligible part of their energy budgets), but because of their remnant heat of formation.  It is easier to detect this remnant heat when the exoplanets are young, as it fades with time.  For this reason, direct imaging tends to target young ($\lesssim 0.1$ Gyr) stars.  By contrast, hot Jupiters are hot because of the intense flux of stellar irradiation they receive, rather than because of internal heat.

While we are ultimately interested in detecting and characterizing Earth-like exoplanets, current astronomical techniques favor the atmospheric characterization of hot and/or large exoplanets.  As technology advances, astronomy will deliver data on cooler and smaller objects.  In the meantime, hot and/or large exoplanets present a unique opportunity for understanding unfamiliar atmospheric regimes and allow us to broaden the scope of our understanding \citep{burrows14}.  The present review focuses on the atmospheres of hot and/or large exoplanets discovered using transits and the questions regarding their basic properties, which are not the traditional focus of the Earth and planetary science literature.

\subsection{Complementarity and conflict with Solar System studies}

A unique attribute of exoplanet science is its ability to simultaneously attract interest from the Earth atmosphere/climate, Solar System and astronomy/astrophysics communities.  A challenge is to unify the relevant bodies of knowledge within these fields towards understanding exoplanetary atmospheres.  The currently characterizable exoplanets (hot Earths/Neptunes/Jupiters\footnote{These exoplanets have approximately the mass and radius of Earth, Neptune and Jupiter, respectively, but the nomenclature does not imply that their atmospheric compositions, thermal profiles, etc, are the same.} and directly imaged exoplanets) reside in chemical and physical regimes for which there exists no precedent in the Solar System.  Well-established techniques in the Earth and Solar System communities may not carry over in a straightforward manner.  Dynamical mechanisms that maintain the circulations on Solar System planets will occur also on exoplanets, but may lead to differing details of circulation because these mechanisms are operating in environments with different stellar flux, chemical composition, rotation rate, etc.  For example, the $\sim 1$ km s$^{-1}$ wind speeds expected in hot Jupiters---and measured, in one case \citep{snellen10}---may imply that traditional radiative transfer methods, which do not consider the relative motion between parcels of gas, need to be revised \citep{mr10}.  On the other hand, some basic physics of atmospheric dynamics has been established by these communities.  For example, hydrostatic balance does not imply a vertically static atmosphere; rather, it implies that vertical motion is so ponderous that hydrostatic balance is quickly re-established (via sound waves).  We expect this insight to hold on hot Jupiters as well as Earth.

It is not for lack of trying that astronomers cannot obtain datasets to the same level of detail as for a given Solar System object.  After all, despite remarkable advances in our ability to spectrally and temporally resolve them, distant exoplanets remain spatially unresolved point sources; we will address this point further in \S\ref{subsect:limit}.  Astronomy will probably never allow us to know intimate details about a given exoplanetary atmosphere (e.g., via spatially-resolved imaging or in-situ probes), but it can elucidate broad trends for dozens and perhaps hundreds of case studies, a point we will address further in \S\ref{subsect:one_vs_many}.

\subsection{Structure of the present review}

The present review will start by summarizing the various astronomical observables from exoplanetary atmospheres (\S\ref{sect:obs}), as motivation for addressing a series of key questions.  These questions are divided into two sections: \S\ref{sect:current} describes a series of questions that are well addressed in the existing literature (up to the point of writing), while \S\ref{sect:future} lists questions that are either only beginning to be addressed or are unaddressed.

Our review builds upon existing reviews of the atmospheric dynamics of hot Jupiters and Earth-like exoplanets \citep{smc08,smc10,showman13b}.  It complements a monograph \citep{seager10} and more observationally-oriented reviews \citep{ds09,sd10}.  It also complements reviews focused on the chemistry and spectral signatures of exoplanets \citep{burrows14,burrows14b,madhu14} and on what one can learn about atmospheric dynamics with a dedicated space mission \citep{psw14}.  We do not discuss either the upper atmosphere (thermosphere, exosphere) or atmospheric escape.  We do not discuss the polarization signatures of the atmosphere.

\section{Observations of Exoplanetary Atmospheres}
\label{sect:obs}

\subsection{Transmission and emission spectra}

\begin{figure}[!ht]
\begin{center}
\includegraphics[width=0.7\columnwidth]{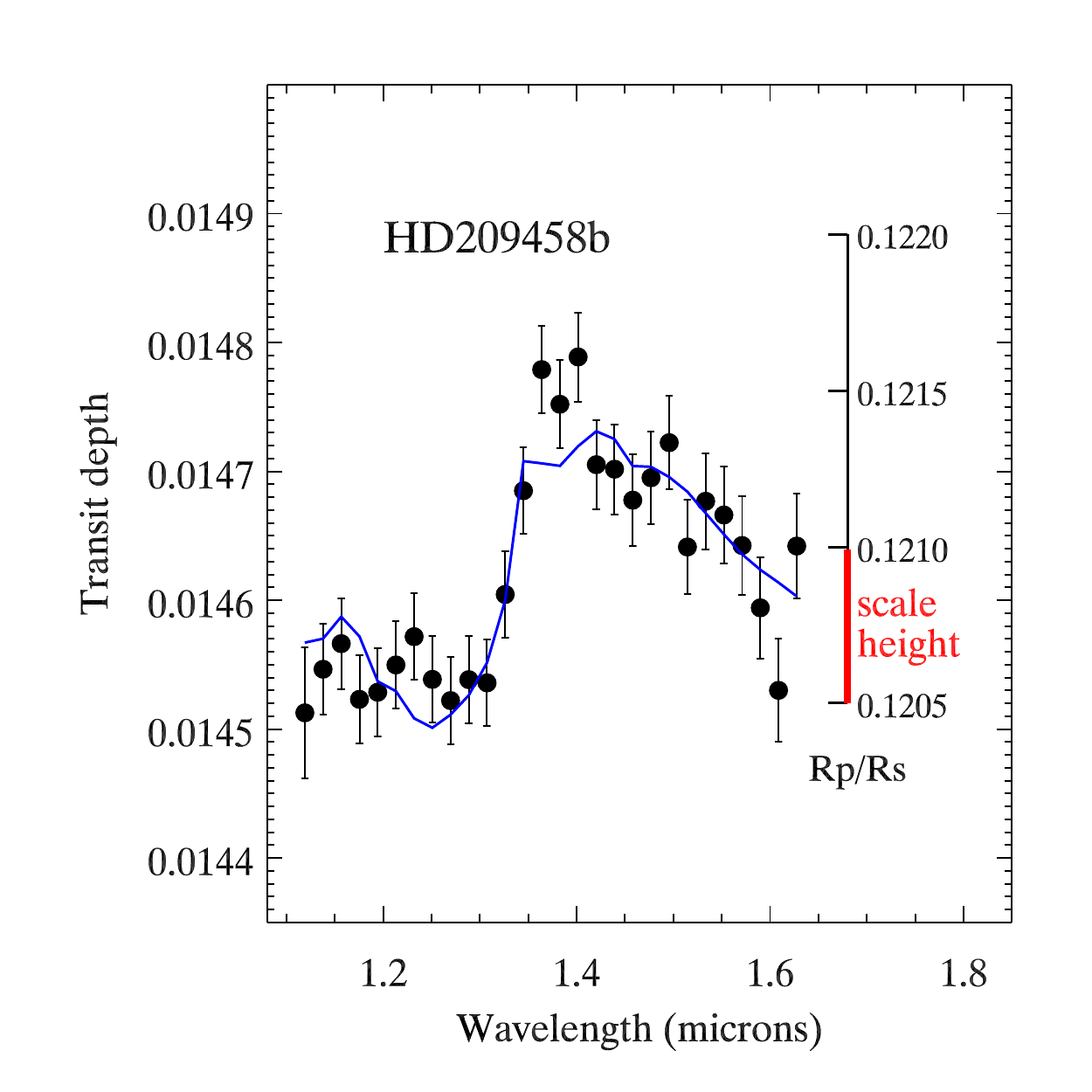}
\end{center}
\caption{Example of a transmission spectrum of the hot Jupiter HD 209458b showing a clean detection of water.  The detected feature would be stronger if the atmosphere was cloud-free, suggesting that clouds are present \citep{deming13}.  Courtesy of Drake Deming.}
\label{fig:trans_spectra}
\end{figure}

Measuring spectra of exoplanetary atmospheres started with obtaining transits and eclipses of hot Jupiters in several broad-band channels (e.g., using the \textit{Spitzer Space Telescope} at 3.6, 4.5, 5.8 and 8.0 $\mu$m) and has evolved to include denser wavelength coverage.  Over time, these techniques have been applied to smaller exoplanets, the so-called hot Neptunes (exoplanets of typically 3--5 Earth radii) and super Earths (exoplanets of typically 1--3 Earth radii), although the latter in particular will continue to provide a major observational challenge for the foreseeable future.  In this subsection, we briefly summarize these techniques, recent results, and show how dynamical constraints arise from them.

Transmission spectra show how the effective radius of a close-in exoplanet varies with wavelength (as seen during transit, when it is backlit by starlight) and therefore provide information on how the atmospheric absorption and scattering vary with wavelength.  At wavelengths where the atmosphere is opaque, stellar radiation directed towards Earth is absorbed or scattered and the exoplanet appears bigger while seen in transit (Figure \ref{fig:schematic0}).  At wavelengths where the atmosphere is transparent, the radiation passes through the atmosphere and the exoplanet appears smaller.  From the early days of exoplanet science, there has been hope that this method will yield the atmospheric composition of exoplanets \citep{ss00,brown01}.  The method has been most spectacularly and successfully applied to the alkali metals Na and K in medium-resolution spectra \citep{sing11b,sing12,nikolov14}; the widths of these lines set an upper limit on the atmospheric pressure probed, via the weakness of pressure broadening, and may in principle constrain the variation of scale height---and hence temperature---with pressure \citep{huitson12}.  Transmission spectroscopy may also be used to constrain the mass of an exoplanet \citep{dewit13}.  These species are in gas phase at the 1000--2000 K temperatures of typical hot Jupiters and are extremely absorbing at specific absorption lines at visible wavelengths \citep{ss00}.   Confidently detecting the existence of molecules such as water, methane and carbon monoxide requires measurements from 1--5 $\mu$m and has proved more challenging.  For water, this has been recently accomplished using measurements from the Hubble Space Telescope \citep{deming13,mandell13}.  Figure \ref{fig:trans_spectra} shows an example of a transmission spectrum, where water was unambiguously detected.


Curiously, a number of exoplanets show relatively flat transmission spectra.  Clouds tend
to be grey and scatter broadly across a wide range of wavelengths, and thus the existence of clouds or hazes at high altitudes provides a likely explanation for such flat spectra \citep{pont08,sing11,knutson14,kreidberg14}.  The existence of these cloud particles requires vertical mixing to be present and this places constraints on the dynamical mixing rates of the atmosphere \citep{hd13,parmentier13}.


Complementarily, emission spectra provide information on the flux of planetary emission as a function of wavelength, which can be converted to a spectrum of brightness temperature versus wavelength.  Typically, such spectra are obtained by subtracting the spectrum obtained just before/after secondary eclipse, when both the star and the planetary dayside are visible, from the spectrum obtained during secondary eclipse, when only the star is visible (Figure \ref{fig:schematic0}).  Since the exoplanet's dayside is oriented towards Earth immediately before and after secondary eclipse, such spectra are of its dayside.  The structures of such emission spectra are affected not only by the atmospheric composition but also by the vertical temperature profile in the exoplanet's atmosphere.  Spectra exhibiting prominent absorption bands, e.g., from water, suggest atmospheres where temperatures decrease with altitude \citep{char08}, while spectra exhibiting prominent emission bands suggest atmospheres where temperatures increase with altitude (``temperature inversions").  Temperature inversions help maintain thermochemical equilibrium and the higher temperatures also generally mitigate against disequilibrium due to photochemistry \citep{moses11}.

Most observations of secondary-eclipse spectra seem to indicate temperatures decreasing with altitude in the layers probed.  By contrast, several early analyses that suggested such thermal inversions on exoplanets \citep{knutson08} are now being refined, and in some cases the evidence for inversions is weakening or disappearing \citep{hansen14,zellem14}.  Regardless, atmospheric dynamics can affect the mean temperature of the dayside and hence the mean depths of the secondary eclipses, and therefore these spectra contain information about the strength of heat transport from the dayside to the nightside of these planets.  In particular, hot Jupiters receiving unusually high stellar flux seem to have dayside temperatures close to local radiative equilibrium, implying inefficient transport of thermal energy from the dayside to the nightside \citep{ca11b}.  By contrast, hot Jupiters receiving less stellar flux in many cases have dayside brightness temperatures significantly less than the local radiative equilibrium temperature that would prevail in the absence of circulation, implying a more efficient transport of heat from day to night.  Circulation models of hot Jupiters naturally produce this type of behavior \citep{php12,ps13}.

\subsection{Phase curves and brightness maps}

\begin{figure}
\begin{center}
\includegraphics[width=0.75\columnwidth]{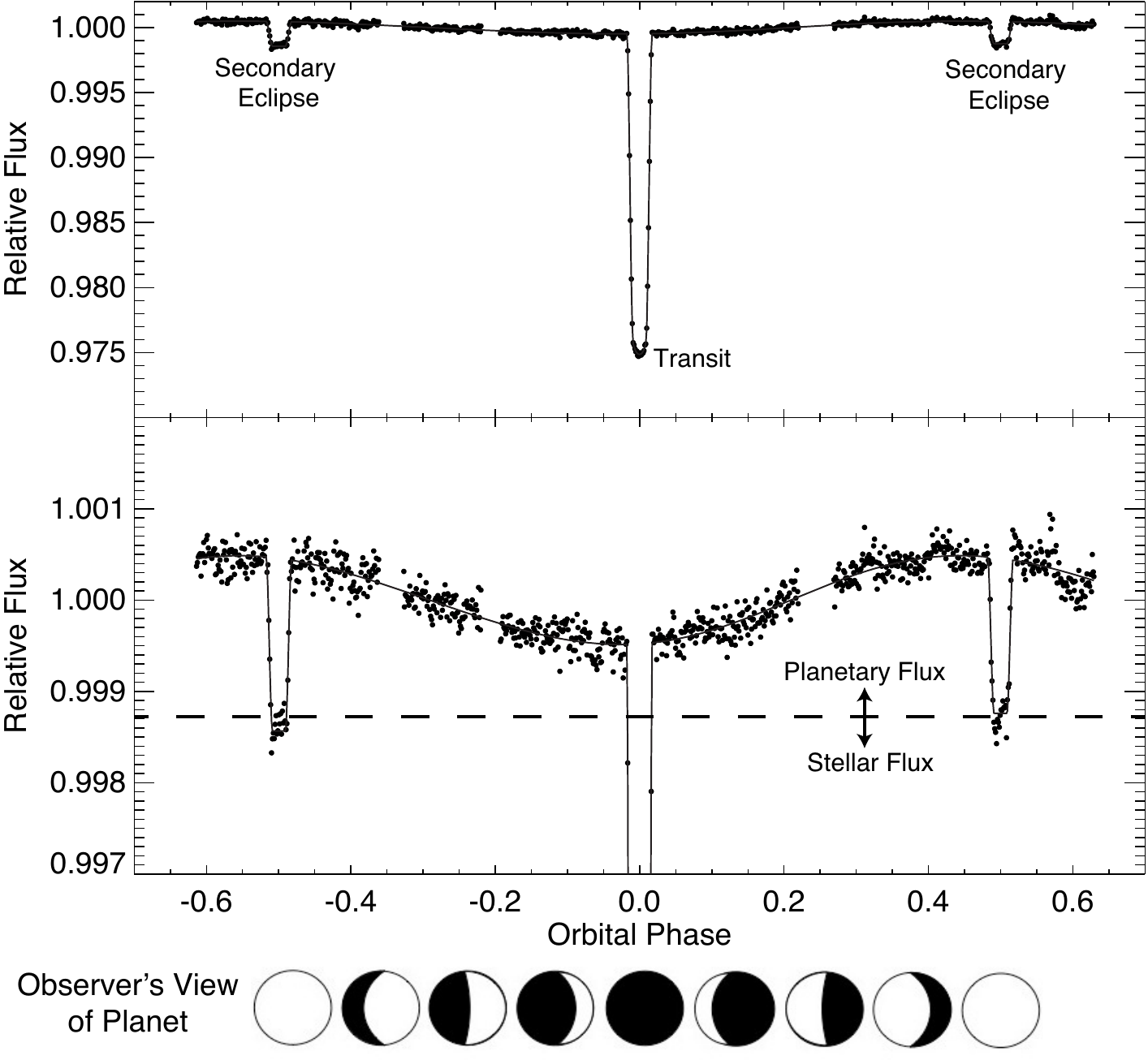}
\includegraphics[width=0.75\columnwidth]{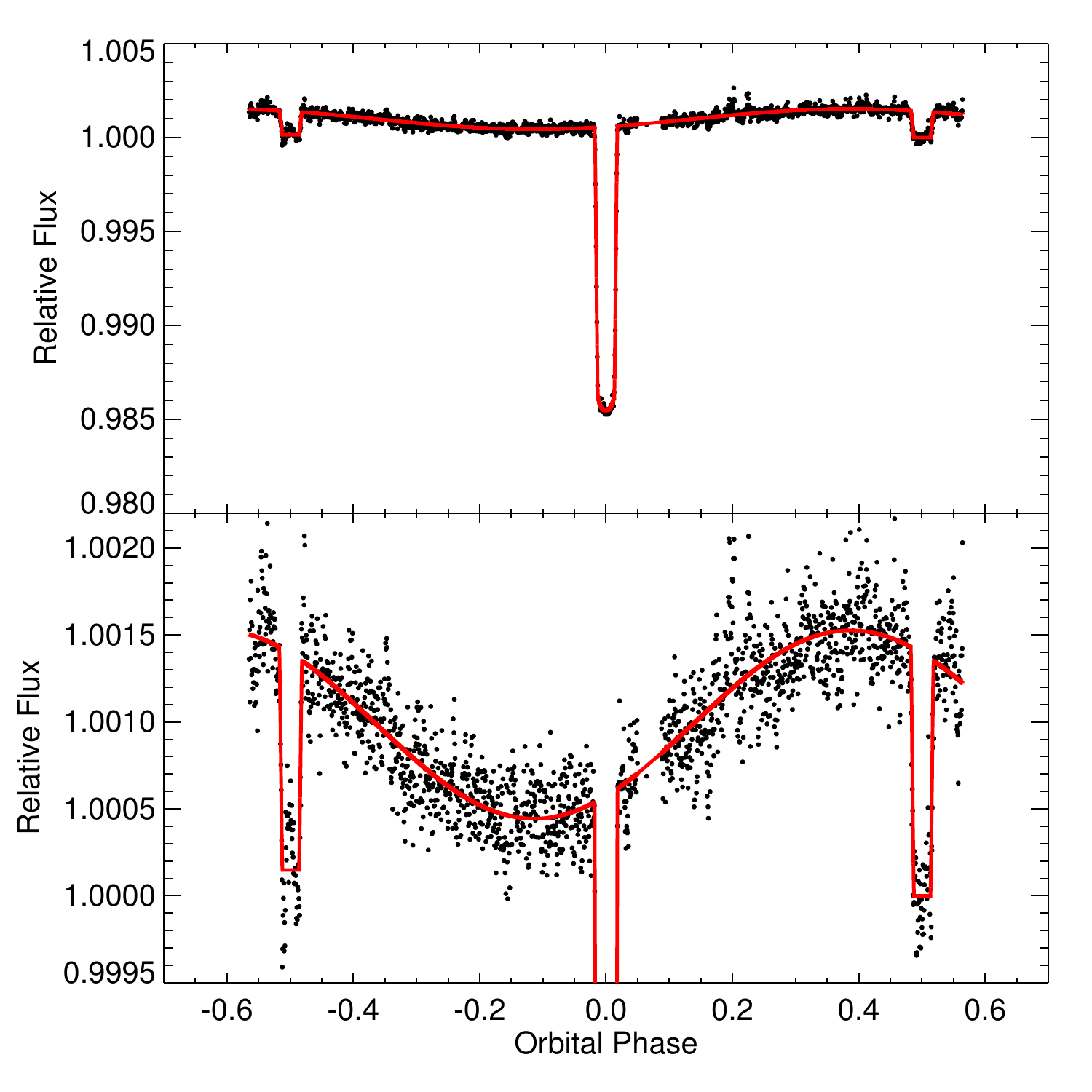}
\end{center}
\caption{Infrared phase curves of the hot Jupiters HD 189733b (top panel) and HD 209458b (bottom panel).  Courtesy of Heather Knutson \citep{knutson07,knutson09a,knutson12} and Robert Zellem \citep{zellem14}.}
\label{fig:phase_curve}
\end{figure}

\begin{figure}
\begin{center}
\includegraphics[width=\columnwidth]{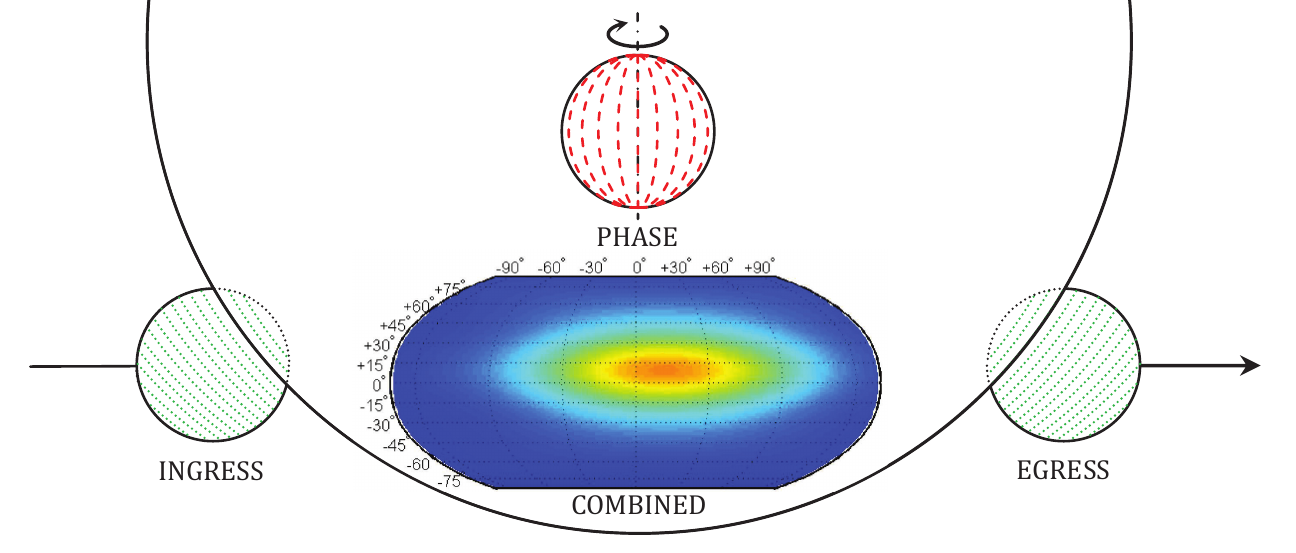}
\end{center}
\caption{Combined eclipse and phase map of the hot Jupiter HD 189733b.  Courtesy of Julien de Wit \citep{dewit12}.}
\label{fig:eclipse_map}
\end{figure}

Often overlooked in favor of transmission or emission spectra---partly due to the time-consuming nature of obtaining them---phase curves contain a wealth of data that is multi-dimensional in nature \citep{harrington06,knutson07}.  A phase curve records the rise and ebb of flux at different orbital phases of the exoplanet (Figure \ref{fig:schematic0}).  It is the longitudinal distribution of flux from the exoplanetary atmosphere convolved with its geometric projection to the observer \citep{ca08}.  Deconvolving the phase curve yields a 1D ``brightness map", the most iconic example of which is shown in Figure \ref{fig:phase_curve} for the hot Jupiter HD 189733b \citep{knutson07}.  Infrared phase curves measured at different wavelengths probe different depths or pressure levels of the atmosphere, allowing for the construction of 3D maps.

In special cases where the orbital planes of the exoplanet and its star do not coincide, high-cadence measurements before and after the secondary eclipse yield complementary, latitudinal (north-south) information on the exoplanetary atmosphere (Figure \ref{fig:eclipse_map}).  This technique is known as ``eclipse mapping".  When combined with the phase curve, which contains purely longitudinal (east-west) information, a 2D brightness map may be constructed, the first example of which was reported for the hot Jupiter HD 189733b \citep{dewit12,majeau12,majeau12b}.

High-quality light curves can be inverted to obtain the latitudinal-mean brightness temperature versus longitude \citep{ca08}, a quantity that provides significant information about the atmospheric dynamics.  In particular, exoplanets whose dynamics have equalized the temperatures at all longitudes will exhibit flat lightcurves, whereas those with order-unity day-night temperature differences will observe a high-amplitude phase variation throughout the orbit, comparable to the depth of the secondary eclipse.  Observations indicate that cooler planets such as HD 189733b have modest phase-curve variations, suggesting modest brightness temperature variations from dayside to nightside, whereas highly-irradiated exoplanets have large phase variations, suggesting fractional brightness temperature differences from dayside to nightside of order unity \citep{ps13}.  This behavior complements similar findings from secondary-eclipse spectra alone \citep{ca11b} and has been qualitatively explained with dynamical models \citep{php12,ps13}.

Moreover, the \textit{offsets} of the flux minima and maxima in the lightcurves provide information about the longitudinal displacement of the hottest and coldest regions from the substellar and antistellar points, respectively.  In Figure \ref{fig:phase_curve}, for example, one can see that the flux extrema \textit{precede} the secondary eclipse and transit, indicating that the hot and cold regions are displaced to the east.  First observed for HD 189733b \citep{knutson07,knutson09a,knutson12}, an eastward-shifted hot spot has also been detected for a variety of other hot Jupiters, including $\upsilon$ Andromedae b \citep{crossfield10}, HAT-P-2b \citep{lewis13} and HD 209458b \citep{zellem14}.  Interestingly, this feature was predicted in 3D circulation models some years before its discovery \citep{sg02}, and has now been reproduced in a wide variety of GCMs \citep{cs05,dd08,showman09,rm10,hmp11,hfp11,php12,mayne13a}.  Section \ref{sect:current} will follow up these theoretical results in more detail.

Optical phase curves record the reflectivity of the atmosphere across longitude and set constraints on the size and relative abundances of condensates or aerosols \citep{demory13,hd13}.  For the very hottest Jupiters, their thermal emission extends into the optical range of wavelengths---their phase curve thus contains both reflected light and thermal emission.  Examples include the optical phase curves of HAT-P-7b by the \textit{Kepler Space Telescope} \citep{borucki09} and CoRoT-1b by \textit{CoRoT} \citep{snellen09}.

For rare cases of exoplanets that reside very close to their stars, such that they become gravitationally distorted, regular flux modulations known as ``ellipsoidal variations" are produced in the phase curve \citep{welsh10,cowan12}.  In the case of Kepler-76b, relativistic beaming has been detected \citep{faigler13}.

\subsection{Ultra-high resolution cross-correlation techniques}

\begin{figure}
\begin{center}
\includegraphics[width=0.7\columnwidth]{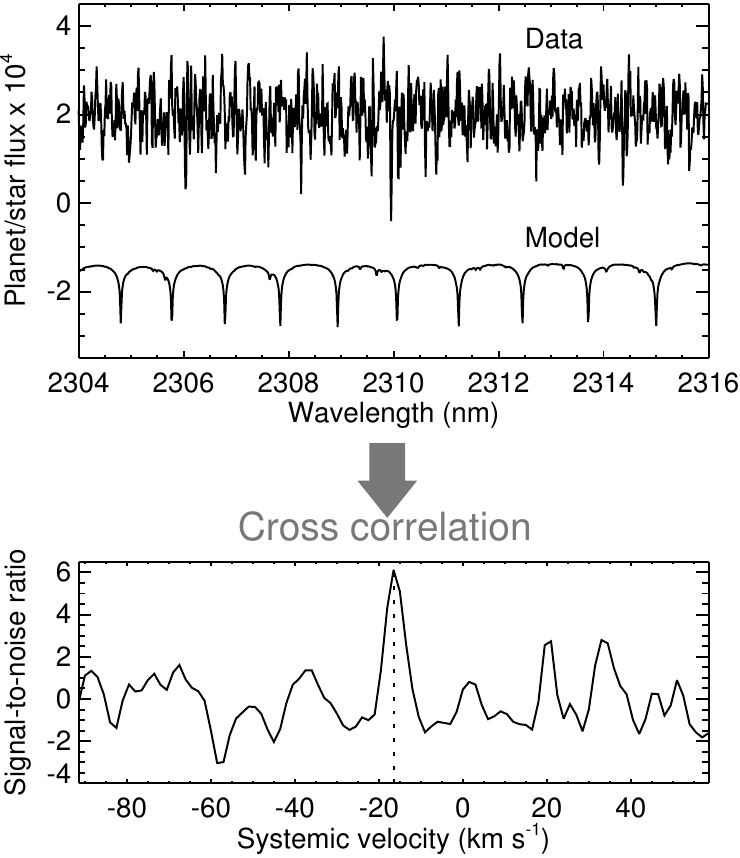}
\end{center}
\caption{Example of ultra-high resolution spectroscopy for studying exoplanet atmospheres. The top panel shows 452 spectra of the system $\tau$ Bo\"{o}tis observed with CRIRES at the VLT, after removing telluric and stellar lines, and contains the absorption signatures of various molecules as well as noise. For clarity, only 1/4 of the covered spectral range is shown.  The spectra are shifted to the rest frame of the exoplanet $\tau$ Bo\"{o}tis b and co-added. The expected exoplanet spectrum, containing molecular absorption from carbon monoxide, is shown by the model plotted in the same panel.  After cross-correlating the data with the model, the exoplanet signal appears at a S/N of about 6, at the known systemic velocity of $\tau$ Bo\"{o}tis ($-16.4$ km s$^{-1}$).  Courtesy of Matteo Brogi and Ignas Snellen.}
\vspace{0.1in}
\label{fig:ultrahigh}
\end{figure}

A complementary technique to using space-based telescopes is the use of ultra-high-resolution (spectral resolution $\sim 10^5$) spectrographs, mounted on 8 meter-class telescopes, to record transmission spectra, albeit over a narrow wavelength range (Figure \ref{fig:ultrahigh}).  The measured spectra are cross-correlated with a theoretical template of a specific molecule to determine (or rule out) its presence.  For example, 56 absorption lines of carbon monoxide were analyzed and an overall blueshift of $2 \pm 1$ km s$^{-1}$ was reported relative to the expected motion of the hot Jupiter HD 209458b, which was interpreted to be a signature of global atmospheric winds \citep{snellen10}.  (Such an analysis also allows for the orbital motion of the exoplanet to be directly measured, leading to an accurate estimation of its mass.)  The interpretation of these measurements have been challenged as being an artefact of a mildly eccentric orbit \citep{montalto11}, due to a misunderstanding of the orbital properties being measured, and subsequent work has demonstrated that the orbital motion of HD 209458b produces a blueshift of $\sim 0.1$ km s$^{-1}$, insufficient to solely account for the reported blueshift \citep{crossfield12,showman13a}.

The same method has been used to detect carbon monoxide in the \textit{non-transiting} hot Jupiters $\tau$ Bo\"{o}tis b \citep{brogi12,rodler12} and HD 179949b \citep{brogi14}, which allowed their orbital inclinations to be measured, and water \citep{birkby13} and carbon monoxide \citep{dekok13} in HD 189733b.  The ability to characterize non-transiting exoplanets greatly increases the potential sample size of exoplanets to be studied.  Improvements in the data quality will eventually allow for the procurement of phase curves.

\subsection{The Inflated Hot Jupiter Problem}

\begin{figure}
\begin{center}
\includegraphics[width=0.8\columnwidth]{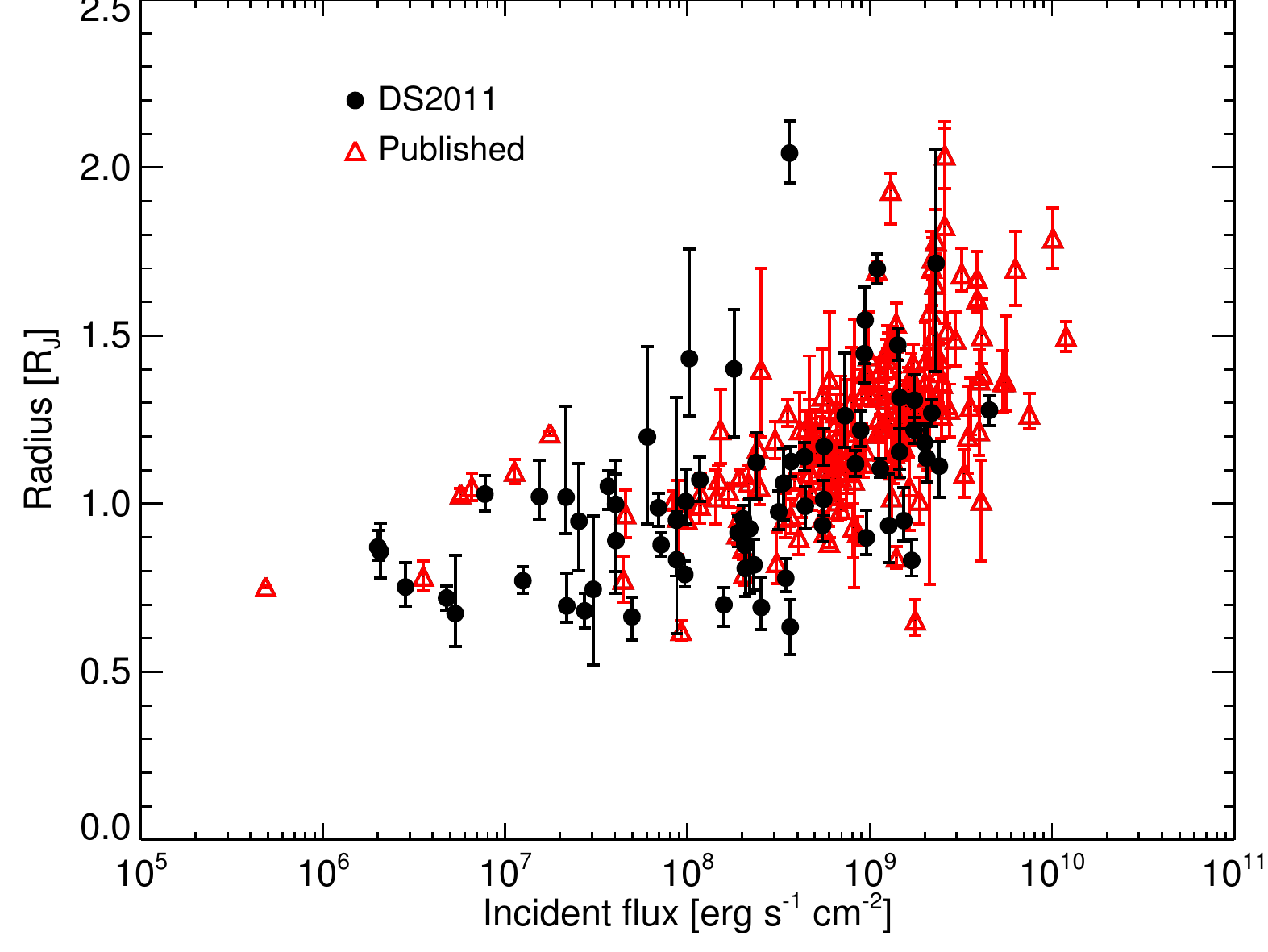}
\end{center}
\caption{Measured radii of hot Jupiters versus the incident stellar flux upon them.  Hot Jupiters appear to be more ``inflated" when they are more irradiated.  For comparison, the solar constant at Earth is $\sim 10^6$ erg cm$^{-2}$ s$^{-1}$.  Courtesy of Brice-Olivier Demory \citep{ds11}.}
\vspace{0.1in}
\label{fig:inflated}
\end{figure}

A puzzling, observed phenomenon is that hot Jupiters appear to have larger radii when they are more irradiated, with some objects having radii twice that of Jupiter's \citep{baraffe10} (Figure \ref{fig:inflated}).  Keeping hot Jupiters ``inflated" is non-trivial, since Jupiter-mass objects are expected to be partially degenerate\footnote{Matter becomes ``degenerate" when it is packed so closely together that its pressure no longer depends on temperature and is determined by quantum mechanical effects.}.  A plausible mechanism for inflation has to deposit a sufficient amount of energy deep within the interior of the exoplanet.  Proposed mechanisms include the dynamical deposition of heat via vertical mixing \citep{gs02,sg02}, Ohmic dissipation \citep{bs10,pmr10a,pmr10b} and semi-convection \citep{cb07}, although we note that semi-convection does not depend on the strength of stellar irradiation.  Both vertical mixing and Ohmic dissipation are mechanisms that are intimately related to atmospheric dynamics.  Furthermore, the location of the radiative-convective boundary in hot Jupiters is influenced by atmospheric dynamics, which enhances the long-term contraction and exacerbates the inflated hot Jupiter problem \citep{rs14b}.

\subsection{Albedos and albedo spectra}

Optical measurements of the secondary eclipse directly yield the geometric albedo---the albedo at zero phase angle---of either the atmosphere or surface or an exoplanet \citep{seager10}, provided the exoplanet is not so hot that its thermal emission leaks into the optical range of wavelengths \citep{hd13}.  When corrected for contamination by thermal emission, surveys of hot Jupiters reveal that the geometric albedo is uncorrelated, or weakly correlated at best, with the incident stellar flux \citep{hd13}, stellar metallicity and various properties (surface gravity, mass, radius and density) of the exoplanets \citep{anger14}.  Curiously, the geometric albedos of super Earths appear to be statistically higher, as a population, compared to hot Jupiters \citep{demory14}, implying either that their atmospheric properties are fundamentally different or that these albedos might be associated with their surfaces.

Measuring both the dayside and nightside brightness temperatures, in the infrared, allow for the albedo and the circulation efficiency to be simultaneously constrained; applied to a sample of hot Jupiters, this has demonstrated that a range of albedos and circulation efficiencies exist \citep{ca11a}.

In the case of the hot Jupiter HD 189733b, its relatively bright star allows for the albedo spectrum of the exoplanet to be measured and demonstrates that its atmosphere is more reflective at shorter/bluer optical wavelengths \citep{evans13}.  The measured albedo spectrum is consistent with an atmosphere dominated by Rayleigh scattering (either by condensates or hydrogen molecules) and absorption by sodium atoms \citep{hd13} or unusually small condensates \citep{hml14}.

\subsection{Temporal variability}

For spatially unresolved exoplanets, measurements of the temporal variability of their atmospheres offer an unprecedented opportunity to probe their atmospheric dynamics.  As an example, if Earth were an exoplanet, measuring its temporal power spectrum would reveal a broad peak at 30 to 60 days associated with the Madden-Julian Oscillation \citep{po92}.  To date, temporal variability has not been detected for any exoplanetary atmosphere.  An upper limit of 2.7\% has been set on the dayside variability of HD 189733b \citep{agol10,knutson12}, thus ruling out 2D shallow water simulations that neglect thermal forcing and predict $\sim 10\%$ variability \citep{cho03,cho08}, although it remains consistent with the $\sim 1\%$ predictions of 3D simulations \citep{showman09}.  Attempts have been made to detect temporal variability from HD 149026b \citep{knutson09b}, HD 209458b \citep{crossfield12} and $\upsilon$ Andromedae b \citep{crossfield10}.

\section{Current Questions Concerning the Dynamics of Exoplanetary Atmospheres}
\label{sect:current}

\begin{figure}
\begin{center}
\includegraphics[width=0.9\columnwidth]{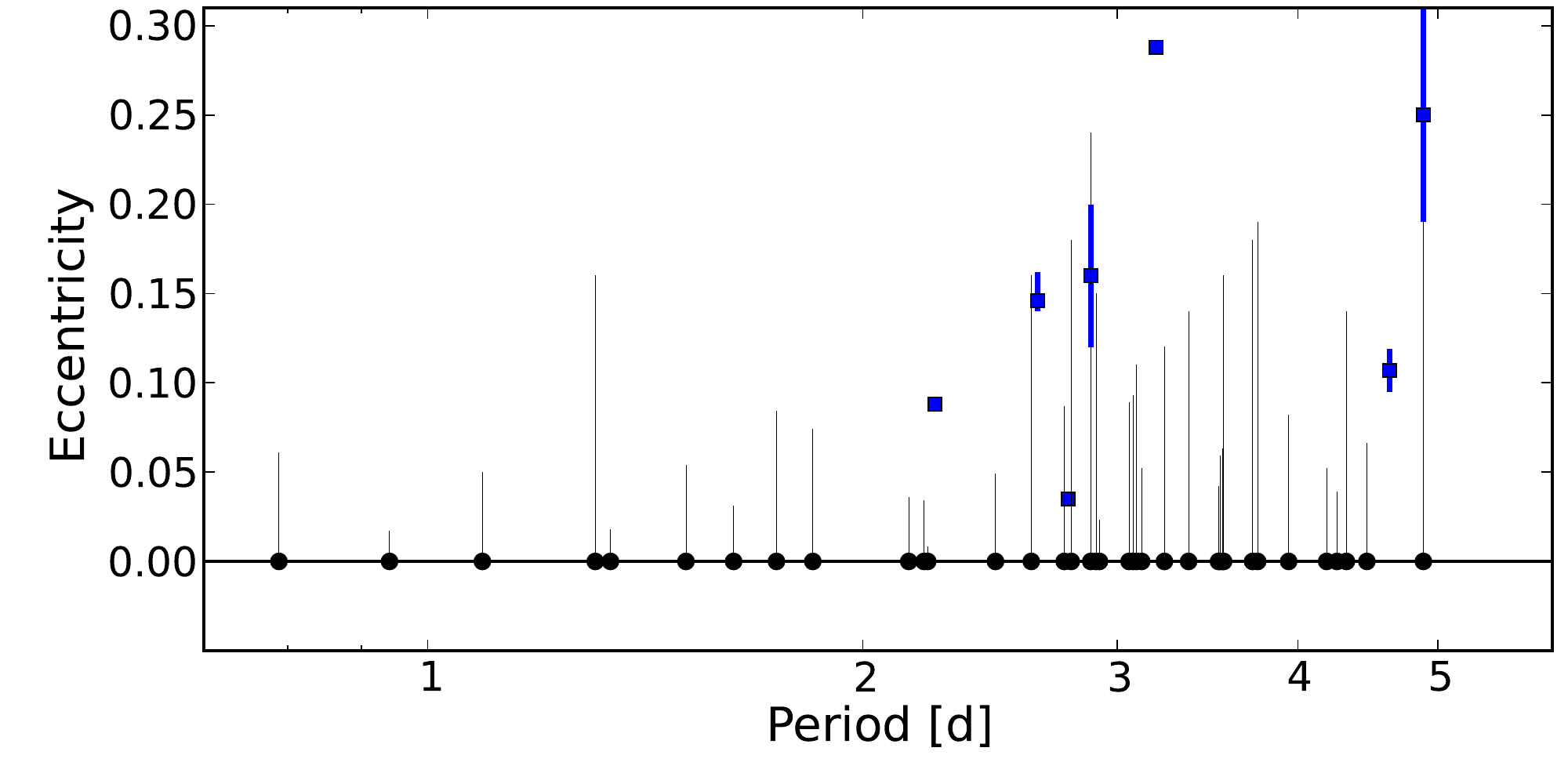}
\end{center}
\caption{Orbital eccentricity of a sample of exoplanets, with measured masses and radii, as a function of the orbital period (in units of Earth days) \citep{pont11}.  A Markov Chain Monte Carlo (MCMC) analysis is used to correct for the effect that the orbital eccentricity is bounded from below by zero, in order to obtain more accurate estimations of the eccentricities and their uncertainties.  Data points consistent with zero eccentricity are represented by the black circles, while the blue squares are measurements of non-zero eccentricity.  Exoplanets that are located close enough to their host stars (i.e., with orbital periods less than about 2 days) appear to reside on circular orbits.  Courtesy of Nawal Husnoo and Fr\'{e}d\'{e}ric Pont \citep{pont11}.}
\vspace{0.1in}
\label{fig:pont11}
\end{figure}

The rotation rate of an exoplanet is a fundamental property that affects all aspects of the atmospheric dynamics and therefore the thermal structure.  It is an important input in general circulation models.  Generally, this quantity is not easily obtained from astronomical observations.  However, for hot Jupiters/Neptunes/Earths residing on close-in, circular orbits, the orbital and rotation periods are expected to be equal; the orbital period is easily measured using the transit and radial velocity techniques.  

The lowest-energy orbital state of an exoplanet occurs when it resides on a spin-synchronized, circular orbit.  The time scale for spin synchronization is \citep{blm01},
\begin{equation}
t_{\rm syn} = \frac{ 8 Q \Omega M a^6}{45 G M_\star^2 R^3},
\label{eq:sync}
\end{equation}
where $\Omega$ is the rotation rate, $M$ is the mass of the exoplanet, $G$ is Newton's gravitational constant and $M_\star$ is the stellar mass.  The time scale for circularization takes a different form \citep{gs66},
\begin{equation}
t_{\rm circ} = \frac{ 4 Q M a^{13/2}}{63 G^{1/2} M_\star^{3/2} R^5}.
\label{eq:circ}
\end{equation}
The ``tidal quality factor" $Q$ is the reciprocal of the fraction of tidal energy dissipated per orbit.  Its value has been experimentally estimated to be $Q \sim 10$--100 for Earth \citep{knopoff64}.  For Jupiter, it is estimated that $Q \sim 10^4$--$10^5$ \citep{lainey10}.  Generally, since $t_{\rm circ}$ has a steeper dependence on $a$ than $t_{\rm syn}$, it is expected that circularized orbits are also spin-synchronized.  (The converse is not true: tidally de-spun but eccentric orbits may exist, where the simple permanent dayside and nightside picture does not hold.)  Astronomical measurements of the orbital eccentricity of hot Jupiters show that their orbits are consistent with being circularized for orbital periods of $\lesssim 2$ days (Figure \ref{fig:pont11}) \citep{pont11}.  Thus, for hot Jupiters at least, the rotation period may be indirectly inferred.

Brown dwarfs typically have rotation periods of about 1 to 12 hours \citep{rb08} and the first observational estimate of rotation for a directly imaged exoplanet shows that its rotation period is likewise short---close to 8 hours \citep{snellen14}.  These rapid rotations place brown dwarfs and directly imaged gas giants in a different dynamical regime than close-in exoplanets that are synchronously---and therefore more slowly---rotating \citep{sk13}.

\subsection{What are the basic thermal structures of highly-irradiated atmospheres (as predicted by theory)?  How is heat transported from the permanent dayside to the nightside?}
\label{subsect:thermal}

\begin{figure}
\begin{center}
\vspace{-0.2in}
\includegraphics[width=0.9\columnwidth]{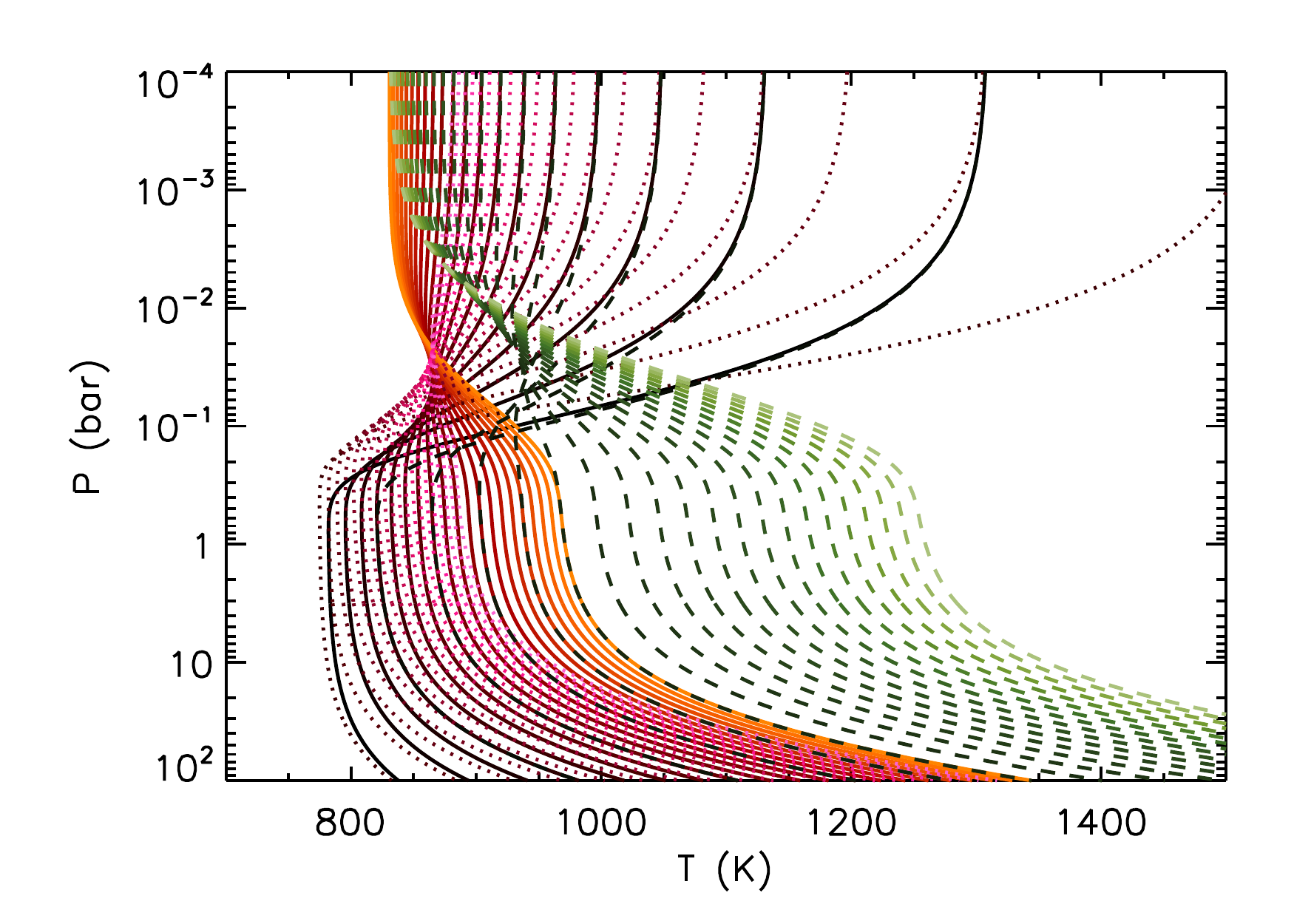}
\end{center}
\caption{Examples of temperature-pressure profiles.  The families of solid, dotted and dashed curves correspond to model atmospheres with pure absorption, shortwave scattering present and longwave scattering present, respectively.  Each family of curves keeps the shortwave opacity fixed at $\kappa_{\rm S}=0.01$ cm$^2$ g$^{-1}$ and performs a sweep of the longwave opacity ($\kappa_{\rm L}=0.001$--0.02 cm$^2$ g$^{-1}$).  Stellar irradiation determines the thermal structure down to $\sim 10$ bar, where internal heat takes over.  For illustration, we have adopted a surface gravity of $g = 10^3$ cm s$^{-1}$, an equilibrium temperature of $T_{\rm eq} = 850$ K and an internal temperature of $T_{\rm int}=200$ K.  Models are taken from \cite{guillot10}, \cite{hhps12} and \cite{hml14}.}
\vspace{0.1in}
\label{fig:tp}
\end{figure}

The zeroth order way of describing an atmosphere is to state its ``equilibrium temperature" (assuming a vanishing albedo): $T_{\rm eq} = T_\star (R_\star/2a)^{1/2}$, where $T_\star$ is the effective stellar temperature, $R_\star$ is the stellar radius and $a$ is the spatial separation between the exoplanet and the star.  The stellar flux incident upon the exoplanet is $2 \sigma_{\rm SB} T^4_{\rm eq}$ (the ``stellar constant"), where $\sigma_{\rm SB}$ is the Stefan-Boltzmann constant.  Favored by astronomers, the stellar constant is an atmosphere-independent quantity, which is convenient to compute and state.  Unsurprisingly, a survey of the Solar System planets and moons reveals that such a description is inadequate for describing the actual atmospheric temperatures.  Additionally, there is theoretical interest to understand the thermal structures of exoplanetary atmospheres both vertically and horizontally, as observations are starting to constrain these quantities.

The temperature on exoplanets in general, and especially on synchronously-rotating exoplanets with permanent daysides and nightsides, will depend significantly on latitude and longitude.  For tidally-locked exoplanets, intense stellar irradiation on the permanent dayside produces a horizontal temperature gradient, which drives zonal and meridional winds.  The extent to which these winds redistribute heat from the dayside to the nightside, and the resulting day-night temperature contrast, is one of the outstanding questions in current exoplanet science.  Several authors have suggested that whether the day-night temperature difference is small or large can be understood by a comparison of the characteristic timescale over which air advects horizontally from day to night, $t_{\rm adv}$, with the characteristic timescale over which air gains or loses energy by absorption of starlight and radiation of infrared energy, $t_{\rm rad}$ \citep{sg02,cs05,showman08,hmp11,hfp11,ca11b,php12}.  According to this scenario, the fractional day-night temperature difference is small when $t_{\rm rad} \gg t_{\rm adv}$ and large when $t_{\rm rad} \ll t_{\rm adv}$.  Under conditions typical of hot Jupiters, both timescales span the range $\sim10^4$--$10^5$ s, suggesting that hot Jupiters reside in the regime where day-night temperature differences may be large (unlike on Earth).  Comparisons
of these two timescales in 3D numerical simulations of hot Jupiters, with each timescale evaluated a posteriori from the simulation results, suggests that the transition from small to large fractional day-night temperature difference does occur approximately where $t_{\rm rad} \sim t_{\rm adv}$ \citep{php12}.  Moreover, it was predicted that when $t_{\rm rad} \sim t_{\rm adv}$, the winds can distort the temperature pattern, leading to hot spots that are displaced from the substellar point \citep{sg02}.  In particular, they predicted a fast, broad eastward jet stream at the equator (see \S\ref{subsect:circulation}), which causes a displacement of the hottest regions to the east.  As summarized in \S\ref{sect:obs}, this phenomenon has now been observed on several hot Jupiters and been reproduced in a wide range of 3D circulation models.

Nevertheless, this timescale comparison is not predictive, as $t_{\rm adv}$---which is simply a manner of expressing the wind speeds---is unknown a priori and depends on many atmospheric parameters \citep{ps13}.  Unfortunately, it is therefore not possible to predictively evaluate the criterion; one can only estimate after the fact whether it is consistent with a given numerical simulation. Moreover, this timescale comparison neglects a role for other important timescales in the problem, including those for planetary rotation, friction, vertical advection, and propagation of various types of waves both horizontally and vertically.  These almost certainly affect various aspects of the dynamics, including the day-night temperature differences.   Using an idealized, analytical model for the wind speeds and fractional day-night temperature differences, it was demonstrated that, under certain conditions, the $t_{\rm rad}$ versus $t_{\rm adv}$ comparison breaks down \citep{ps13}.  This occurs under conditions of weak day-night forcing, when the horizontal advection term is small relative to vertical advection in the thermodynamic energy equation.   Generally, a comparison between the \textit{vertical} advection timescale and the radiative timescale provides a better description of the regime transition between small and large fractional day-night temperature differences \citep{ps13}.  Furthermore, the advection timescales can be represented in terms of the timescales for planetary rotation, friction, wave propagation and radiation \citep{ps13}.

A crucial question concerns whether the mean temperature profile, on both the dayside and nightside, decreases or increases with height; as described in Section \ref{sect:obs}, this question is directly amenable to observational characterization using secondary eclipses and lightcurves at a variety of wavelengths.  Simple atmospheric theory \citep{guillot10,pierrehumbert} shows that, in the absence of scattering, one expects temperature to decrease with altitude near photospheric levels when the visible opacity is less than the infrared opacity, where temperature increases with height when the visible opacity exceeds the infrared opacity (Figure \ref{fig:tp}).    On hot Jupiters, the presence of a visible absorber can lead to such a ``stratosphere" confined to the dayside alone, with the nightside temperature decreasing strongly with altitude \citep{showman09,hfp11}.  

If the advection timescale is less than the chemical timescale of a given transition (e.g., carbon monoxide to methane), then chemical disequilibrium may be induced by atmospheric dynamics \citep{cs06,burrows10}.  In the case of HD 189733b, examining its color-magnitude diagram at different orbital phases of the exoplanet reveals both a temperature and chemical transition from the dayside to the nightside \citep{triaud14} and allows the data to begin to address this question.

\subsection{What are the global circulation structures of highly-irradiated, tidally-locked exoplanets?}
\label{subsect:circulation}

\begin{figure}[!h]
\begin{center}
\subfigure{\hsize.46\columnwidth\includegraphics[width=0.5\columnwidth]{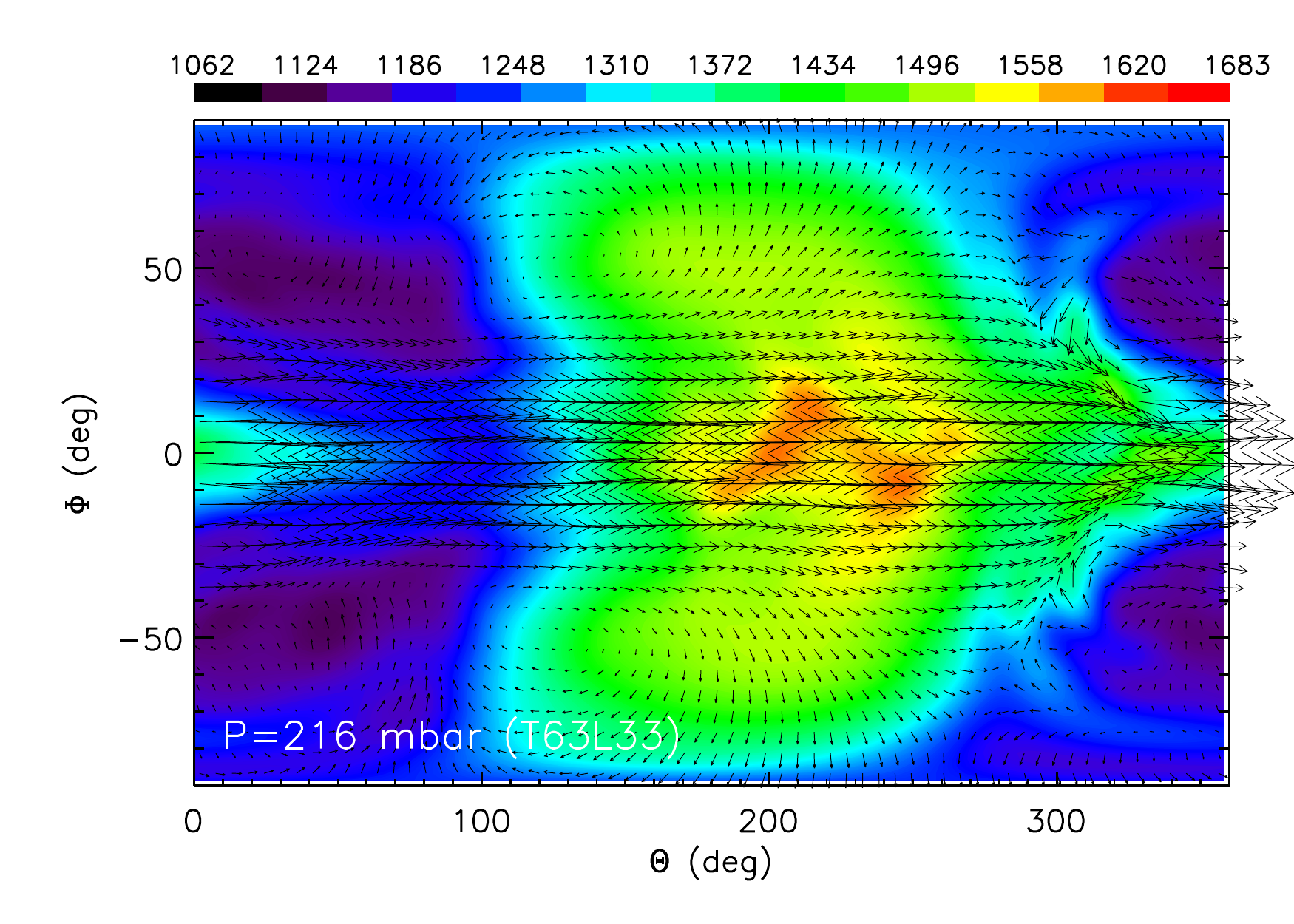}}
\subfigure{\hsize.46\columnwidth\includegraphics[width=0.48\columnwidth]{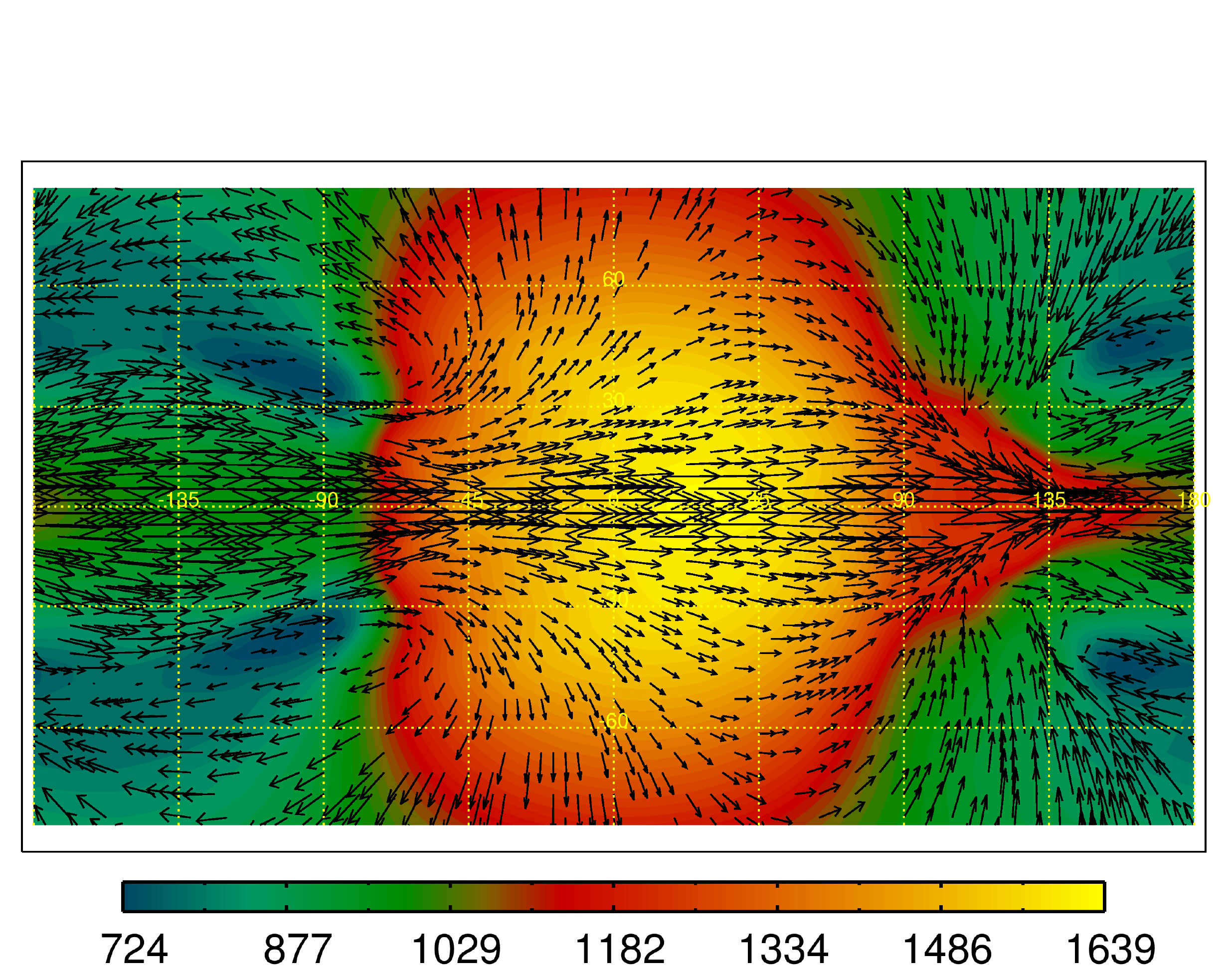}}
\subfigure{\hsize.46\columnwidth\includegraphics[width=0.45\columnwidth]{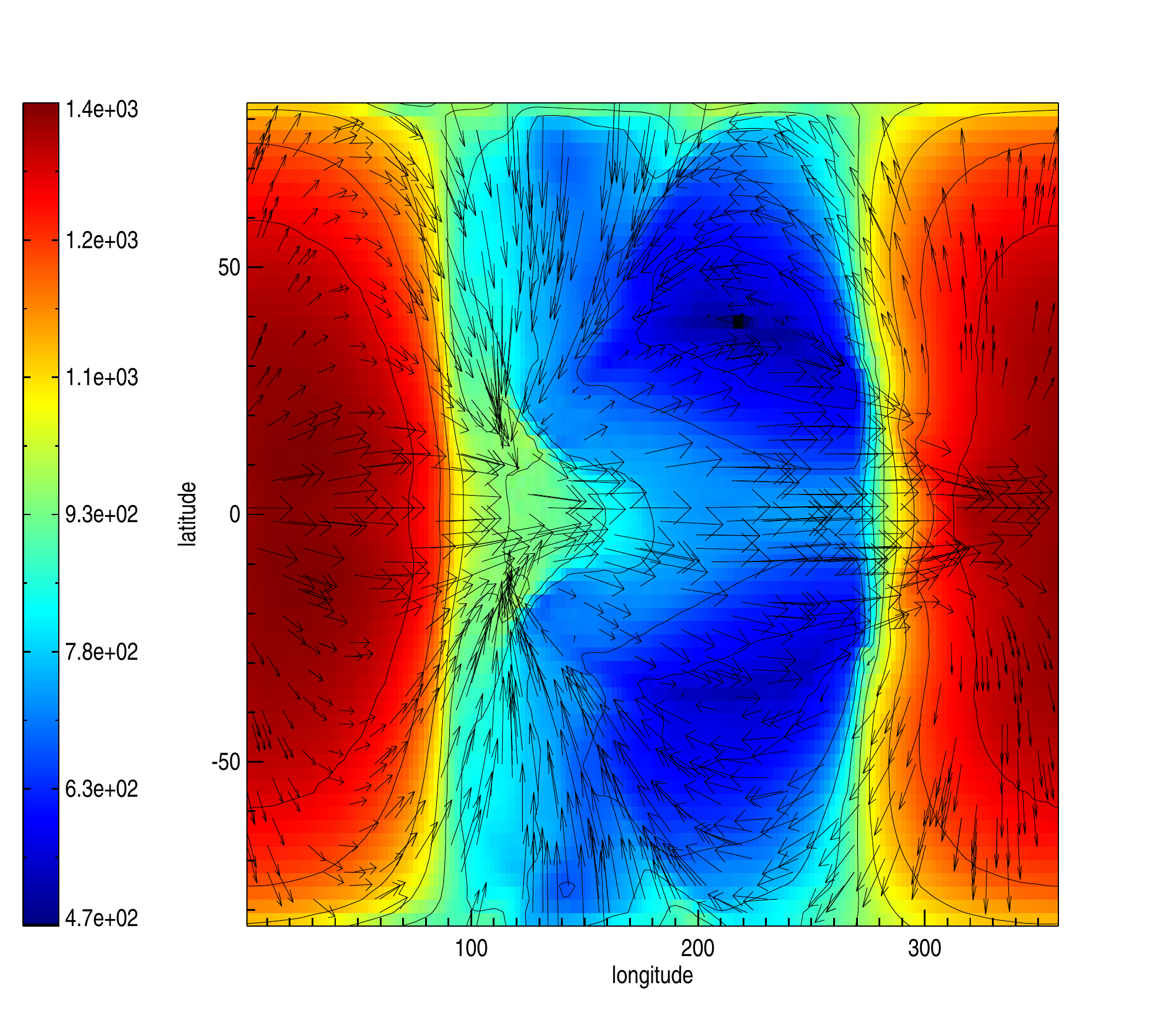}}
\subfigure{\hsize.46\columnwidth\includegraphics[width=0.5\columnwidth]{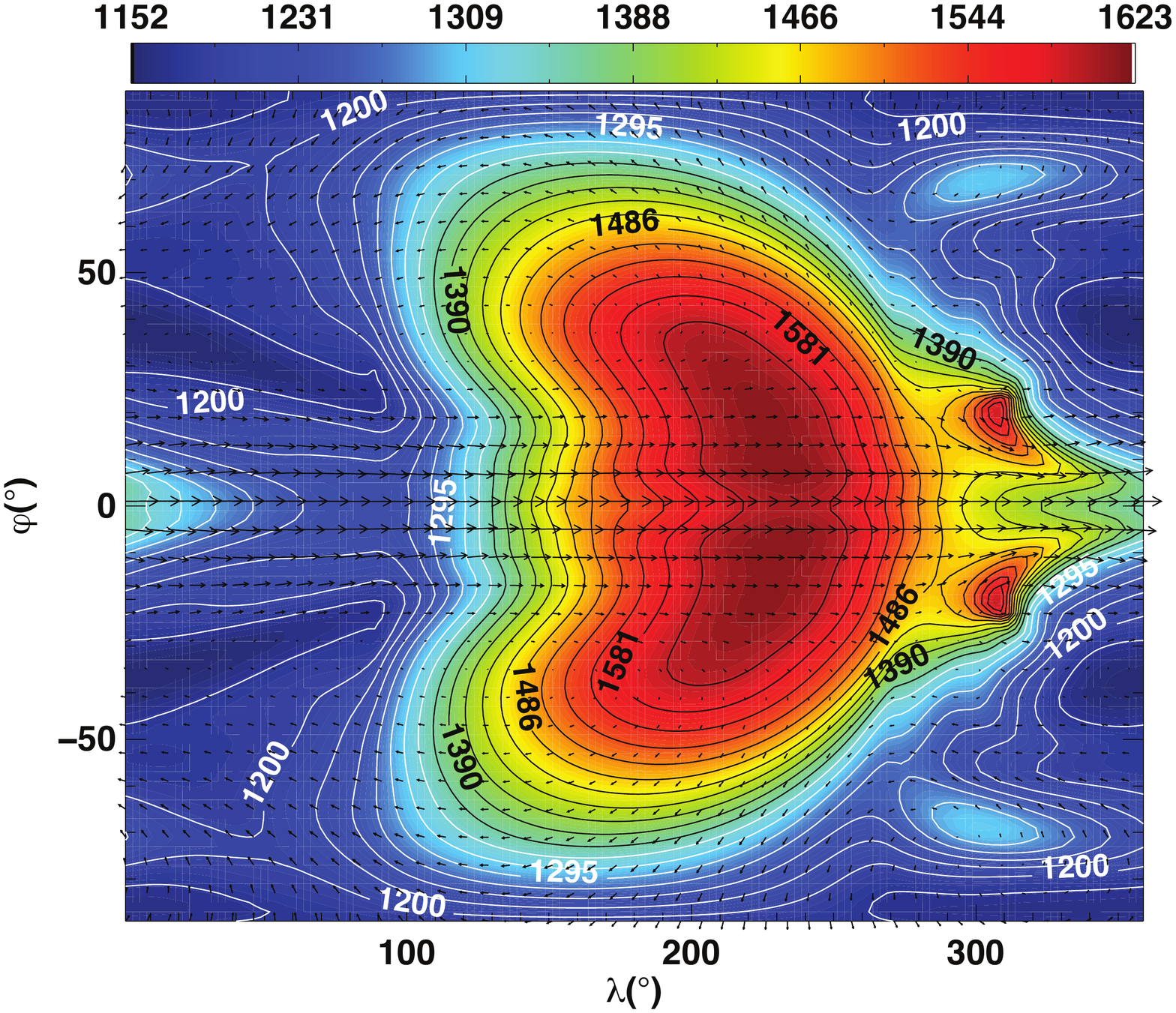}}
\subfigure{\hsize.46\columnwidth\includegraphics[width=0.55\columnwidth]{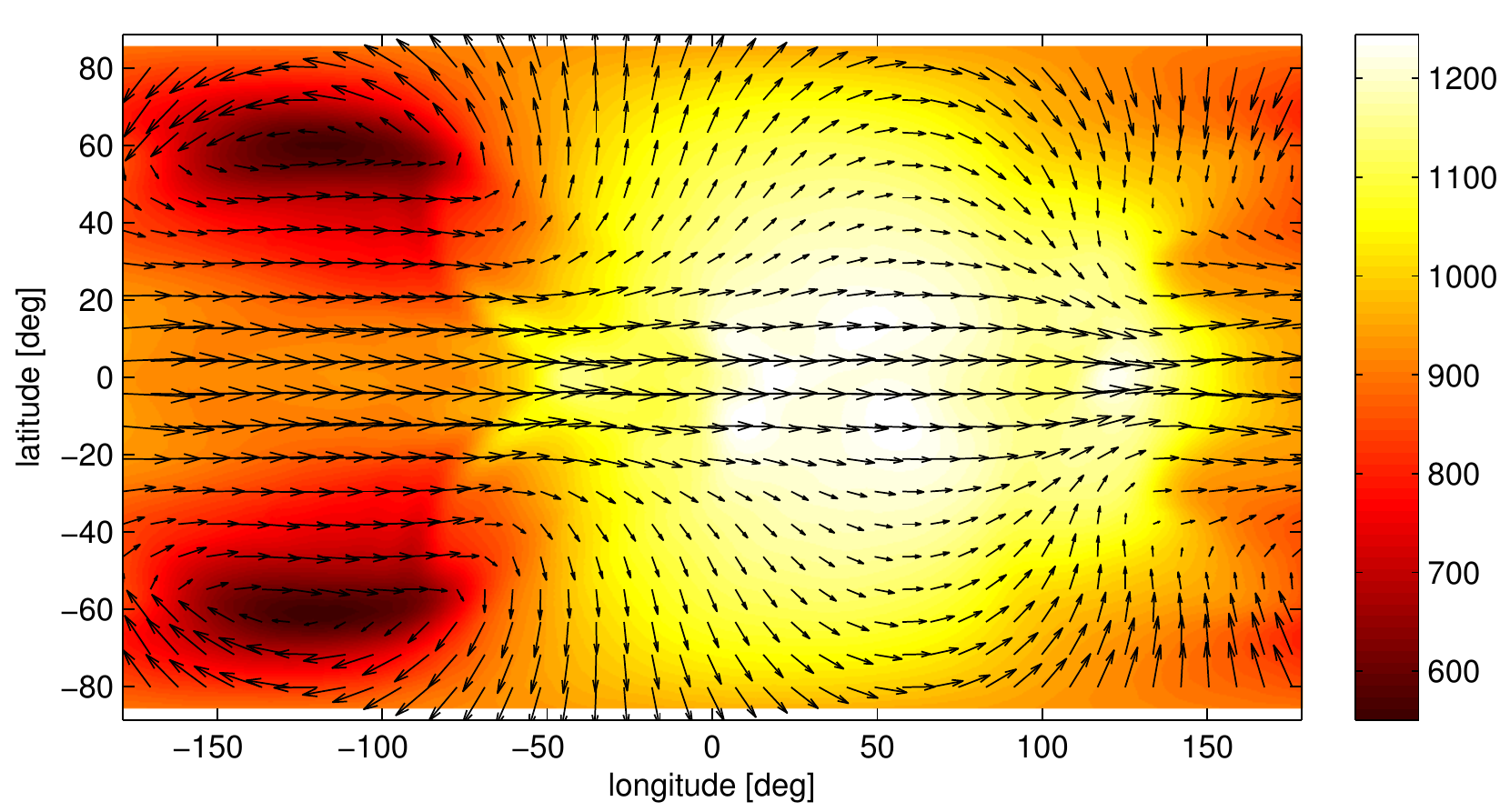}}
\subfigure{\hsize.46\columnwidth\includegraphics[width=0.42\columnwidth]{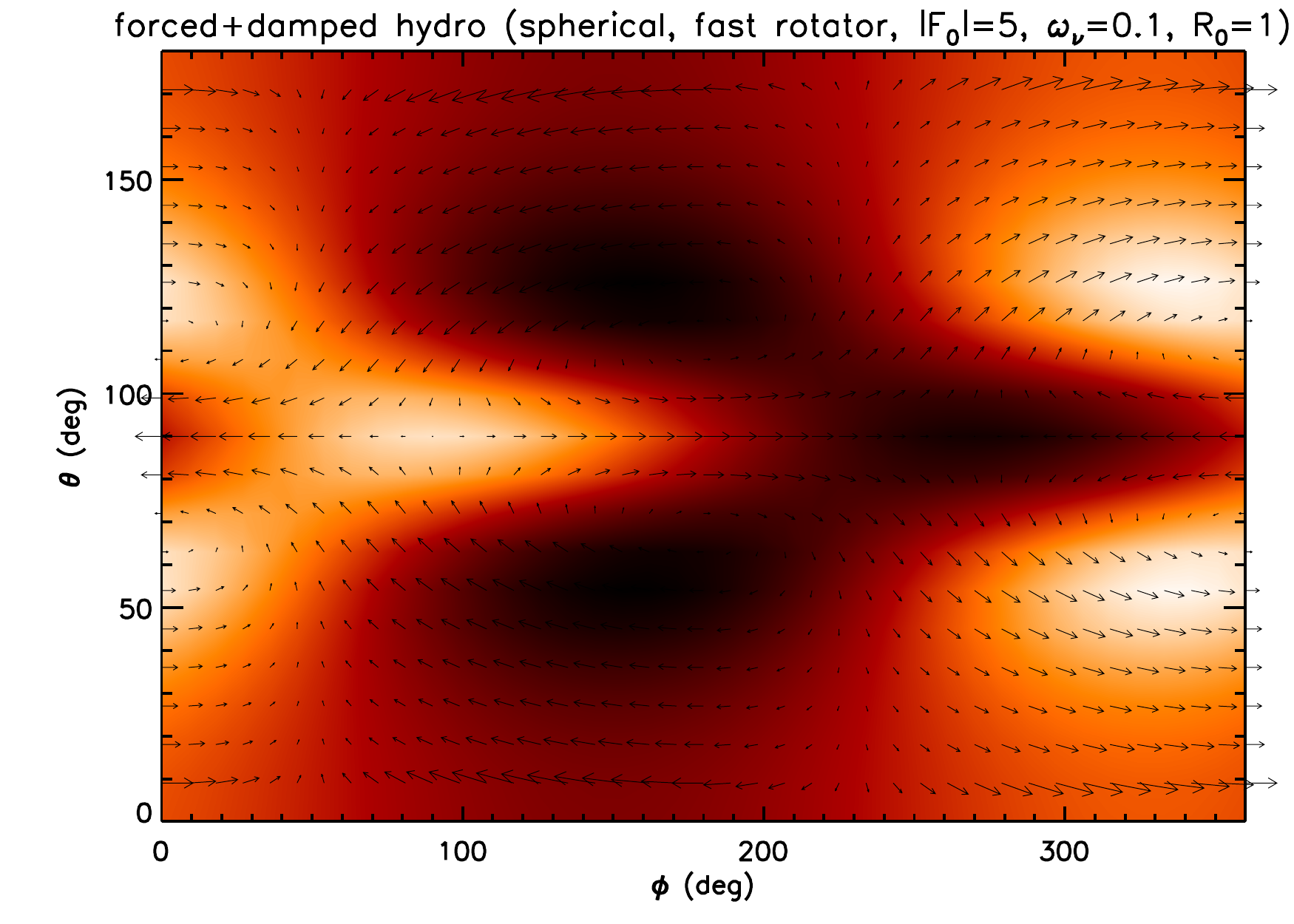}}
\end{center}
\caption{Examples of calculations from general circulation models of hot Jupiters.  Despite the use of different input parameters, techniques to treat radiation and numerical schemes for atmospheric dynamics, the chevron-shaped feature appears to be a generic outcome of the hot exoplanet regime.  Courtesy of Kevin Heng (top left panel; using the \texttt{FMS} GCM), Emily Rauscher (top right panel; using the \texttt{IGCM}), Ian Dobbs-Dixon (middle left panel; using a customized code), Nathan Mayne (middle right panel; using the U.K. Met Office GCM) and Adam Showman (bottom left panel; using the \texttt{MITgcm}).  The bottom right panel shows an analytical model from \cite{hw14}, generalized from the work of \cite{matsuno66}, \cite{gill80} and \cite{sp10,sp11}.}
\label{fig:gcm}
\end{figure}

The dominant dynamical feature emerging from 3D circulation models of hot Jupiters is the existence of a fast, broad eastward-flowing equatorial jet at and near the photosphere (Figure \ref{fig:gcm}) \citep{sg02,cs05,cs06,dd08,dd10,dd12,showman08,showman09,lewis10,hmp11,hfp11,php12,rm10,rm12b}.  This so-called ``equatorial superrotation" is interesting, because such an eastward equatorial jet corresponds to a local maximum of angular momentum per unit mass with respect to the planetary rotation axis; waves or eddies are required to maintain such a feature.  (For example, the atmospheres of Titan and Venus superrotate.)  In many cases, the jet is accompanied by an obvious chevron-shaped feature, with a preferential tendency for northwest-southeast tilts in the northern hemisphere and southwest-northeast tilts in the southern hemisphere.  Analytical theory and idealized numerical models suggest that the equatorial superrotation results from standing, planetary-scale Rossby and Kelvin wave modes that result from the strong day-night thermal forcing \citep{sp10,sp11,tsai14}.  The resulting ``Matsuno-Gill" pattern \citep{matsuno66,gill80} also causes the chevron pattern in at least some cases \citep{hw14}.  The jet width is controlled by the equatorial Rossby deformation radius \citep{sp11},
\begin{equation}
{\cal R}_o \sim \left( \frac{N H R}{2 \Omega} \right)^{1/2},
\end{equation}
where $N$ is the Brunt-V\"{a}is\"{a}l\"{a} frequency, $H$ is the scale height, $R$ is the planetary radius and $\Omega$ is the rotation rate.  Thus, faster rotation leads to a narrower equatorial jet, and in some cases the emergence of additional eastward jets at high latitudes \citep{showman08,showman09,kataria13}.  The circulation in 3D models is also accompanied by strong mean-meridional circulation cells; the strength and depth of these cells are sensitive to the stellar irradiation, becoming stronger and deeper for more irradiated atmospheres \citep{php12}.  The equatorial jet strengthens as the metallicity\footnote{In astronomy, one refers to the elements heavier than hydrogen and helium as ``metals".  Atmospheres with higher metallicities effectively have higher mean molecular weights and thus smaller scale heights.} increases \citep{lewis10}.

Despite the overall tendency toward equatorial superrotation, theory and idealized simulations predict distinct circulation regimes for synchronously-rotating hot Jupiters depending on the incident stellar flux and other parameters.  Using idealized models, it is suggested that when the stellar irradiation is particularly large \citep{showman13a}---perhaps on hot Jupiters with mean temperatures exceeding $\sim 2000$ K---the atmospheric radiative time constant is so short that radiation damps the large-scale Rossby and Kelvin waves that drive superrotation, leading to an atmosphere dominated at low pressures by day-night flow.  On the other hand, when the radiative time constant is very long, the day-night thermal forcing gradient becomes less important, and the mean equator-pole heating gradient instead should become dominant; such a circulation would exhibit a circulation with strong zonal (east-west) banding.    These two dynamical regimes have very different predictions for the wind behavior as observed with high-resolution spectroscopy during transit \citep{kr12,showman13a}.  They also lead to a transition from large to small fractional day-night temperature differences \citep{php12,showman13a,ps13}.  Hot Jupiters like HD 189733b and HD 209458b lie at an intermediate point along this continuum.  Nevertheless, magnetic effects may significantly influence the dynamics for particularly hot exoplanets (see \S\ref{subsect:mhd}), a phenomenon whose effects are still being worked out.

The basic global circulation structure appears to be robust to the presence of a non-zero orbital eccentricity \citep{kataria13}.  A key difference is that spin-synchronized exoplanets on eccentric orbits are ``flash-heated" at periastron, compared to the constant stellar flux received by exoplanets on circular orbits.  If the radiative time scale is sufficiently long, this flash heating remains imprinted for several rotational periods and manifests itself as ``ringing" in the infrared phase curves \citep{ca11a,kataria13}.  To date, this ringing has not been detected.

The dynamical mechanisms that maintain the circulation in Solar System atmospheres such as those of Earth and Jupiter differ in detail from those operating in hot Jupiters, but they share a similar foundation.  On most Solar System planets (with Mars being an exception), the solar day is less than the atmospheric radiative time constant, such that (at least deep in the atmosphere) longitudinal day-night thermal forcing is subdominant relative to the equator-to-pole thermal forcing.    On Earth, the poleward heat transport occurs via the Hadley circulation at low latitudes and baroclinic instabilities at high latitudes \citep{vallis06}.  Coriolis forces in the poleward-flowing upper branch of the Hadley circulation lead to so-called subtropical jets near the poleward edges of the Hadley cells.  Baroclinic instabilities cause radiation of Rossby waves in the mid-latitudes, and the propagation and breaking of these waves drive the eddy-driven jet streams in the mid-latitudes.  Similar mechanisms may be important in maintaining the zonal jets on Jupiter and Saturn. In slowly rotating atmospheres such as those of Venus and Titan, by contrast, the Hadley circulation is nearly global, with the subtropical jets at high latitudes, and equatorial superrotation emerges from instabilities that transport angular momentum to the equator.   In contrast to these solar system examples, the day-night forcing seems to play the overriding role for the typical hot Jupiter.  This forcing induces global-scale waves, including Rossby waves, that cause equatorial superrotation and help maintain the overall thermal structure.  Despite the differences in detail, all of these atmospheres share fundamental similarities in the importance of heat and angular momentum transports in shaping the circulation, and in the importance of wave-mean-flow interactions in controlling the structure of the jet streams.  

\subsection{Is there a continuum of thermal and circulation structures and dynamical mechanisms?  Do we expect the atmospheres of highly-irradiated exoplanets to be fundamentally different from those of the Solar System objects or brown dwarfs?}

An argument that is often made is: if we do not understand the atmospheres of Earth and the Solar System bodies, what chance do we have of understanding the atmospheres of exoplanets?  The answer is that it depends on the question one is asking.  In Earth climate studies, one is interested in predicting temperature variations on fine scales: a warming of 2 K is significant for terrestrial climate change, but of little consequence for interpreting the global properties of an exoplanetary atmosphere.

Despite sharing the term ``Jupiter" in their names, Jupiter and hot Jupiters have little in common, dynamically, besides both being gas giants.  Jupiter is a fast rotator, while hot Jupiters are expected to be tidally locked and slow rotators.  Internal heat and solar irradiation are comparable for Jupiter.  In hot Jupiters, the stellar irradiation flux is $\sim 10^4$ times stronger than if their internal heat fluxes were to be comparable to Jupiter's and the tidal forces experienced are $\sim 10^6$ times stronger.  Based on estimations of the Rossby and Rhines numbers, Jupiter is expected to possess small vortices and a large number of narrow zonal jets, whereas the vortices and jets on hot Jupiters are expected to be global in scale.  Atmospheric circulation in hot Jupiters is believed to be both strong and deep, implying that measured properties may be representative of the entire atmosphere (and not just confined to the cloud-top level as for Jupiter).  Since the temperatures on Jupiter are $\sim 100$ K, while those on hot Jupiters are $\sim 1000$--3000 K, there is little reason to believe that their atmospheric chemistries and aerosol properties will resemble each other, although this remains an active topic of research.

On the other hand, Jupiter might bear more resemblance to brown dwarfs: substellar objects that are massive enough to ignite deuterium and lithium burning within their cores, but not massive enough to sustain full-blown nuclear fusion like in stars.  Dynamically, brown dwarfs and Jupiter may be similar if rotation is fast \citep{sk13}.  Chemically, since brown dwarfs cover a temperature range $\sim 100$--3000 K (termed the sequence of Y, T and L dwarfs), they exhibit a continuum of spectral signatures that may link smoothly to Jupiter's \citep{burrows03}.  Characterizing the coolest Y dwarfs is a work in progress.  There are currently no empirical constraints on the atmospheric dynamics of directly imaged exoplanets, but the analysis of spectra suggests that their chemical and aerosol properties are similar to brown dwarfs \citep{barman11}.

It is interesting to note a source of confusion arising from a misnomer.  In stars, there exists an atmospheric layer known as the ``radiative zone", where energy is transported outwards by radiative diffusion rather than by convection.  It typically sits below a convective zone.  In hot Jupiters, the intense stellar irradiation forms a deep radiative zone that sits above a convective zone; the deepness comes from the stellar flux diminishing the vertical temperature gradient and stabilizing it against convection.  However, unlike for stars, atmospheric circulation exists in the radiative zone of hot Jupiters due to stellar heating.

\subsection{What are the effects of magnetic fields on the atmospheric circulation?}
\label{subsect:mhd}

In highly-irradiated atmospheres, the $\sim 1000$--3000 K temperatures allow for collisional ionization to act upon the alkali metals such as sodium and potassium.  This source of free electrons creates a partially-ionized atmosphere.  If the exoplanet possesses a magnetic field, then it will resist the partially-ionized atmosphere being advected across it---a global manifestation of Lenz's law \citep{bs10,pmr10a}.  This magnetic drag slows down the atmospheric winds and converts kinetic energy into heat (Ohmic heating) \citep{pmr10b}, although this effect is expected to saturate \citep{menou12b}.  When magnetic drag becomes significant, the atmosphere may transition from being dominated by zonal flows to substellar-to-antistellar flow \citep{showman13a}, although the conditions under which this transition is expected to occur remains under debate \citep{batygin13}.  Ohmic dissipation is a plausible mechanism for keeping hot Jupiters inflated and is qualitatively consistent with the observational trends \citep{php12,wl13}, but it remains a matter of debate if it works out in the details \citep{rm13,rs14}.  The presence of clouds is expected to further complicate this issue \citep{heng12a}.

Technically, one needs to solve the governing equations of fluid dynamics in conjunction with the magnetic induction equation,
\begin{equation}
\frac{\partial \vec{B}}{\partial t} = \nabla \times \left( \vec{v} \times \vec{B} \right) -  \nabla \times \left( \eta \nabla \times \vec{B} \right),
\label{eq:induction}
\end{equation}
where $\vec{B}$ is the magnetic field strength, $t$ represent time, $\vec{v}$ is the velocity and $\eta$ is the magnetic diffusivity; $\eta$ is expected to vary by orders of magnitude within a highly-irradiated atmosphere.  The term involving $\eta$ behaves like a diffusion term and accounts for departures from ideal magnetohydrodynamics (MHD).  There exists other terms (Hall, ambipolar diffusion) in the magnetic induction equation, but they are believed to be subdominant \citep{pmr10a}.  To date, most published work invokes the ``kinematic approximation"---the simplification that the magnetic field affects the velocity field of the atmosphere, but not vice versa \citep{menou12,pmr10a,pmr10b,php12,rm13}.  This allows for the effects of magnetic drag and Ohmic dissipation to be studied via post-processing of hydrodynamic simulations.  More recent work has started to formally include MHD, albeit with other simplifying assumptions \citep{batygin13,rs14}.  This technical challenge remains open.

\subsection{What are the effects of aerosols/clouds/hazes on the thermal structure and atmospheric circulation (and vice versa)?}

An inversion occurs in temperature-pressure profiles when the optical opacity exceeds the infrared one \citep{hubeny03,hansen08,guillot10}.  When clouds or hazes are present, they complicate this simple description by introducing both greenhouse and anti-greenhouse effects, depending on their relative strength of absorption versus scattering \citep{hhps12}.  Scattering in the optical generally introduces an anti-greenhouse effect---it warms and cools the upper and lower atmosphere, respectively, by shifting the photon deposition depth to higher altitudes \citep{hhps12,hml14}.  Scattering in the infrared always warms the atmosphere \citep{hml14}.  Examples of model temperature-pressure profiles displaying these basic trends are shown in Figure \ref{fig:tp}.

Clouds exert both radiative and dynamical effects on the atmospheric circulation, while their ability to be kept aloft depends on the local conditions set by the circulation itself.  Even when they are negligible by mass, clouds may alter the temperature-pressure profile via scattering and absorption \citep{hd13}.  If they form a non-negligible part of the mass budget of the atmosphere, they may alter the flow if they are sufficiently coupled to it.  The extent of the coupling depends on the size of the cloud particles relative to the local flow conditions and also determines if the particles ``see" the local vertical flows that keep them aloft \citep{spiegel09,hd13,parmentier13}.  Both scaling arguments and 3D simulations of atmospheric circulation suggest that micron-sized particles (or smaller) should be ubiquitous in hot Jupiters \citep{hd13,parmentier13}.  This may help to explain the flat spectra seen in transmission spectra (\S\ref{sect:obs}) and will exert a strong influence on atmospheric chemistry and radiation if the particles can sublimate on the dayside.

Given the expected diversity of flow and temperature conditions throughout the atmosphere of a tidally-locked, highly-irradiated exoplanet, the abundances and sizes of cloud particles are not expected to be uniform.  The rich structure contained within the optical phase curve of Kepler-7b, which probes the abundances and sizes of clouds within its atmosphere, is consistent with this expectation \citep{demory13,hd13}.

\section{Future Questions}
\label{sect:future}

\subsection{What are the structures of atmospheric winds (as probed by observations)?}

Ultra-high-resolution, ground-based transit observations have led to a tentative measurement of wind speeds in one object \citep{snellen10}, but the potential of this technique to measure circulation structures remains largely untapped.  Measurements of wind speeds will help constrain models of magnetic drag and Ohmic dissipation \citep{bs10,pmr10a,pmr10b} and bracket numerical uncertainties \citep{hmp11}.  The Doppler profile of the winds---whether they are blue- or redshifted---indicates the dynamical regime they are in, whether the atmosphere is dominated by zonal winds or substellar-to-antistellar flow \citep{kr12,showman13a}.  Quantifying the circulation regime of an exoplanet places constraints on the importance of various dynamical mechanisms operating within the atmosphere (e.g., magnetic drag).

These efforts will be significantly advanced with the next generation of giant, ground-based telescopes: the European-Extremely Large Telescope (E-ELT) with a 39-meter mirror; the Giant Magellan Telescope (GMT) with a collecting area equivalent to having a 22-meter mirror; and the Thirty Meter Telescope (TMT).

\subsection{Why do only some exoplanets appear to be cloudy/hazy (as inferred from transmission spectra)?  What observations do we need to break the degeneracies associated with aerosols/clouds/hazes and constrain their properties?}

To date, there is no straightforward explanation as to why some exoplanetary atmospheres appear to be cloudy, while others do not.  The presence of clouds renders the interpretation of spectra degenerate, as the chemical abundances inferred are degenerate with the cloud model assumed \citep{burrows11,lhi13,deming13}.  The degeneracy arises because clouds diminish the strength of spectral features, but this may also be caused by reduced abundances.

Breaking the degeneracies associated with aerosols/clouds/hazes requires a coordinated effort of obtaining albedos, phase curves and ultra-high-resolution transit and secondary eclipse spectra for a given exoplanet.  Measuring the albedo and phase curve will only constrain a degenerate combination of the sizes and abundances of the cloud particles, the total optical depth of the cloud, and the altitude and vertical extent of the cloud \citep{hd13}.  If the infrared phase curve is also obtained, it provides a powerful way of diagnosing the relative abundances of the cloud particles across longitude, since one now has information on the thermal profile of the atmosphere and thus its lofting properties due to dynamics.  If the exoplanet possesses a flat transmission spectrum, then ultra-high-resolution spectra may be able to set constraints on the pressure levels of the cloud deck, since the cores of atomic or molecular lines may still be detected given sufficient spectral resolution \citep{pont13,kph14}.  Furthermore, obtaining transmission and emission spectra shortward of 1 $\mu$m and longward of 8 $\mu$m, along with measurements of the mass and radius of the exoplanet, will set constraints on the size, optical depth and composition of the cloud particles \citep{lee14}.

\subsection{What technical advances do we need to make in our simulation techniques?}
\label{subsect:technical}

Exploring the atmospheric dynamics of exoplanets started with adapting general circulation models (GCMs) designed for the study of Earth.  While these GCMs provide a reasonable starting point for initial investigations, they have a number of shortcomings that need to be remedied in order to address several outstanding questions.  Standard, Earth-based GCMs do not include a treatment of shocks, which are expected to exist in highly-irradiated atmospheres and will convert a significant fraction of the kinetic energy into heat \citep{dd08,lg10,heng12c}.  Standard GCMs also do not include the dynamical effects of magnetic fields.  Instead, several studies have added a ``Rayleigh drag" term into the momentum equation ($-\vec{v}/t_{\rm drag}$) to mimic magnetic drag \citep{pmr10a,pmr10b,php12}, even though hydrodynamic and magnetic drag are expected to possess qualitatively different damping behavior \citep{rs14,hw14}.  Given the interest in clouds as motivated by the astronomical observations, there is a need to develop non-Earth-centric cloud schemes for GCMs.

Table 1 lists and summarizes the GCM studies of exoplanetary atmospheres to date, including the governing equations solved (see Appendix \ref{append:equations} for details), the approximations taken and if key properties are being modeled.  GCMs that formally include magnetic fields are starting to emerge, albeit with their own technical limitations \citep{batygin13,rs14}.  Some GCMs do not solve the ``pole problem", where meridians converging at the poles of a sphere lead to a vanishing computational time step \citep{st12}, and thus have 3D but non-global grids \citep{dd08,dd10,dd12,da13}.  Several of the listed studies do not explicitly demonstrate if they are able to reproduce the standard benchmark test for Earth \citep{hs94}.

Ultimately, in order to understand highly-irradiated exoplanetary atmospheres and predict their emergent properties, we need a GCM that does some combination of the following: solves the Navier-Stokes equation in tandem with the magnetic induction equation, employs a numerical scheme that conserves mass, energy and angular momentum simultaneously, treats shocks (or at least allows for the possibility of shocks emerging), performs multi-wavelength radiative transfer and is able to consider the effects of disequilibrium chemistry.

\subsection{Is there a fundamental limit to what we can learn about spatially unresolved (but spectrally and temporally resolved) exoplanets?}
\label{subsect:limit}

Astronomy has come a long way since the first exoplanet detections in the 1990s.  The promise of using transits, eclipses and direct imaging to measure basic, bulk properties of exoplanetary atmospheres has largely been fulfilled.  Mass-radius diagrams of exoplanets allow a rich set of properties to be inferred, from whether an exoplanet is likely to be rocky or possesses an extended atmosphere to whether its radius is anomalously large as expected by standard physics.  Detailed transit spectroscopy (or spectro-photometry) allow for the presence of molecules to be identified, either using space telescopes or from the ground.  The next generation of telescopes will allow for even more detailed inferences to be made, including the global structures and speeds of winds, the rotation period, the vertical structure of chemical abundances and 3D maps over a broad range of pressure levels.  That all of these accomplishments will come without being able to spatially resolve these exoplanets is in itself a remarkable feat.

Nevertheless, the lack of spatial resolution will ultimately introduce inescapable degeneracies into our interpretations of even next-generation data.  At the day-night (and night-day) terminators of exoplanetary atmospheres, a rich variety of conditions are expected to be present and departures from chemical and radiative equilibrium are expected \citep{cs06,rm10,hfp11}.  Transmission spectroscopy probes some global average of these terminator regions, which may not reflect local conditions.  Similarly, the daysides of exoplanetary atmospheres are expected to exhibit a diverse range of circulations, chemistry and temperatures.  Emission spectroscopy again probes some global average of these conditions.  In breaking these degeneracies, GCMs have a key role to play (once they overcome the technical challenges described in \S\ref{subsect:technical}).  By combining a comprehensive data set consisting of transmission and emission spectra, phase curves and eclipse maps obtained at multiple wavelengths, one can begin to iterate with GCMs to obtain self-consistent solutions and study degeneracies.  If the data is good enough, one may even assess the time dependence of 3D structure in exoplanetary atmospheres.

\subsection{What are the pros and cons of investing resources into a few ``benchmark" exoplanets versus spreading it over many exoplanets to study statistical trends?  What may we expect from future instruments/missions?}
\label{subsect:one_vs_many}

A strength of astronomical observations is the ability to measure basic quantities (mass, radius, density, incident stellar flux, albedo, bulk chemistry) for a large sample of exoplanets, much more than for the 8 planets in our Solar System.  By making these measurements for a diverse sample of stars with different ages, metallicities and stellar types, a catalogue of properties may be built up and may yield important clues on how these exoplanets and their atmospheres formed.  Such an approach is orthogonal to collecting and analyzing samples or sending in-situ probes, as is done for the Solar System.  

On the other hand, a decisive way to advance our understanding of exoplanetary atmospheres is to focus on a handful of exoplanets around nearby, bright stars, which serve as benchmarks for testing both observational and theoretical techniques.  While the heavy investment of telescope time in these objects carries a certain amount of risk \citep{knutson14,kreidberg14}, such measurements must be allowed to continue as the rewards will be immense (e.g., identification of molecules in an Earth-like exoplanet) \citep{seager12}. 

Not all exoplanets are equally characterizable---it is no accident that HD 189733b and HD 209458b, two of the most studied exoplanetary atmospheres, orbit nearby, bright stars.  There are three exoplanet detection missions on the horizon designed to greatly increase the sample of exoplanets around bright stars: TESS \citep{ricker14}, CHEOPS \citep{broeg13} and PLATO \citep{rauer14}, all approved by NASA or ESA and to be launched between 2017 and 2024.  The discoveries made by these missions will enable a significantly larger sample of characterizable exoplanets to be constructed across a broader range of stellar types and atmospheric temperatures, which will inspire more detailed questions regarding their atmospheric dynamics and chemistry.

The James Webb Space Telescope (JWST) will be equipped with three instruments covering 0.6 to 29 $\mu$m at spectral resolutions of 100--3000.  While this is insufficient to resolve individual ro-vibrational lines, it will definitively identify line complexes and thus molecules \citep{line13}.  The JWST is capable of obtaining multi-faceted datasets of exoplanetary atmospheres, including transmission and emission spectra and multi-wavelength phase curves.  The hope is that physical effects such as shock heating or Ohmic dissipation may produce unique signatures of their existence when examined collectively across transmission, emission and phase-curve data.  Multi-wavelength phase curves will provide a wealth of data for constraining atmospheric dynamics, as they measure temperatures across longitude and depth \citep{burrows10}.  Combined with eclipse maps, there is the potential to obtain 2D maps at different altitudes within an exoplanetary atmosphere, enabling the data to decisively confront general circulation models.  For the brightest objects, it will be complemented by the E-ELT, GMT and TMT from the ground, where the identification of molecules will be corroborated.  Complementary, ground-based optical data (shortward of 0.6 $\mu$m) will set additional constraints on cloud or aerosol properties in the atmosphere.

\begin{issues}[THE FUTURE]
The coming decade holds the promise of ground-breaking advances in the study of exoplanetary atmospheres and will witness the decisive confrontation of theory and simulation by the observations.
\end{issues}

\section*{ACKNOWLEDGMENTS}
KH acknowledges financial, logistical and secretarial support from the University of Bern and the University of Z\"{u}rich, as well as grants from the Swiss National Science Foundation (SNSF) and the Swiss-based MERAC Foundation and participation in the Swiss-wide PlanetS framework (PI: W. Benz).  We thank Julien de Wit, Brice-Olivier Demory, Drake Deming, Jonathan Fortney, Robert Zellem and Didier Queloz for constructive feedback following their reading of an earlier version of the manuscript.  KH thanks Sara Seager, Scott Tremaine, Didier Queloz, Willy Benz and Dick McCray for encouragement.

\begin{sidewaystable}[anticlockwise]
\vspace{0.5in}
\centering
\label{tab:summary}
\begin{tabular}{llllllll}
\hline\hline
\multicolumn{1}{l}{Study} & \multicolumn{1}{l}{Approx.} & \multicolumn{1}{l}{Global} & \multicolumn{1}{l}{Irradiated$^\dagger$} & \multicolumn{1}{l}{Radiative$^\dagger$} & \multicolumn{1}{l}{Treats} & \multicolumn{1}{l}{Magnetic} & \multicolumn{1}{l}{Passes Earth$^\ddagger$} \\
\multicolumn{1}{l}{(Alphabetical)} & \multicolumn{1}{l}{Used} & \multicolumn{1}{l}{Grid?} & \multicolumn{1}{l}{Atmosphere?} & \multicolumn{1}{l}{Transfer?} & \multicolumn{1}{l}{Shocks?} & \multicolumn{1}{l}{Fields?} & \multicolumn{1}{l}{Benchmark?} \\
\hline
\vspace{2pt}
\cite{batygin13}$^\clubsuit$ & BQ (3D) & Y & Y & N & N & Y & N \\
\cite{bending12} & PE (3D) & Y & Y & N & N & N & Y \\
\cite{burkert05}$^{\clubsuit\diamond}$ & EE (2D) & N & Y & Y & N & N & N \\
\cite{burrows10} & PE (3D) & Y & Y & N & N & N & Y \\
\cite{cho03} & EB (2D) & Y & N & N & N & N & N \\
\cite{cho08} & EB (2D) & Y & N & N & N & N & N \\
\cite{cs05} & PE (3D) & Y & Y & N & N & N & Y \\
\cite{cs06} & PE (3D) & Y & Y & N & N & N & Y \\
\cite{dd08}$^\clubsuit$ & EE (3D) & N & Y & Y & N & N & N \\
\cite{dd10}$^\clubsuit$ & NS (3D) & N & Y & Y & Y & N & N \\
\cite{dd12}$^\clubsuit$ & NS (3D) & N & Y & Y & Y & N & N \\
\cite{da13} & NS (3D) & N & Y & Y & Y & N & N \\
\cite{hmp11} & PE (3D) & Y & Y & N & N & N & Y \\
\cite{hfp11} & PE (3D) & Y & Y & Y & N & N & Y \\
\cite{kataria13} & PE (3D) & Y & Y & Y & N & N & Y \\
\cite{ll08} & EE (2D) & Y & Y & Y & N & N & N \\
\cite{lewis10} & PE (3D) & Y & Y & Y & N & N & Y \\
\cite{lg10} & NS (2D) & N & Y & N & Y & N & N \\
\cite{ls13} & PE (3D) & Y & Y & N & N & N & Y \\
\cite{mayne13a} & EE (3D) & Y & Y & N & N & N & Y \\
\cite{mayne14}$^\heartsuit$ & EE (3D) & Y & Y & N & N & N & Y \\
\cite{mr09} & PE (3D) & Y & Y & N & N & N & Y \\
\cite{menou12} & PE (3D) & Y & Y & Y & N & N & Y \\
\cite{parmentier13} & PE (3D) & Y & Y & Y & N & N & Y \\
\cite{pmr10a} & PE (3D) & Y & Y & N & N & N & Y \\
\cite{pmr10b} & PE (3D) & Y & Y & N & N & N & Y \\
\cite{php12} & PE (3D) & Y & Y & Y & N & N & Y \\
\cite{pc12} & PE (3D) & Y & N & N & N & N & Y \\
\cite{rm10} (2010) & PE (3D) & Y & Y & N & N & N & Y \\
\cite{rm12a} & PE (3D) & Y & Y & N & N & N & Y \\
\cite{rm12b} & PE (3D) & Y & Y & Y & N & N & Y \\
\cite{rm13} & PE (3D) & Y & Y & Y & N & N & Y \\
\cite{rs14} & AN (3D) & Y & Y & N & N & Y & N \\
\cite{sg02} & PE (3D) & Y & Y & N & N & N & Y \\
\cite{showman08} & PE (3D) & Y & Y & N & N & N & Y \\
\cite{showman09} & PE (3D) & Y & Y & Y & N & N & Y \\
\cite{tc10} & PE (3D) & Y & Y & N & N & N & Y \\
\cite{tc11} & PE (3D) & Y & Y & N & N & N & Y \\
\hline
\hline
\end{tabular}\\
\centering
\tiny
$\dagger$: ``Irradiated" refers specifically to whether the model atmosphere is being forced by stellar irradiation, i.e., whether the irradiation is doing work on the atmosphere.  Simulations that are unforced by irradiation are sometimes termed ``adiabatic".  It is possible for the effects of radiation in irradiated atmospheres to be mimicked without explicitly performing radiative transfer, by adopting a Newtonian relaxation or cooling term in the thermodynamic equation.\\
$\ddagger$: only marked ``Y" if there is either an explicit demonstration in the publication or a clear citation to previous publications describing that the simulation code used is able to reproduce the Held-Suarez benchmark test for Earth \citep{hs94}.  Since it is a 3D test, 2D simulations, by definition, are unable to reproduce it.\\
$\clubsuit$: Employs flux-limited diffusion in the region encompassing the photosphere, an approximation that is strictly valid only in optically thin or thick situations.  \\
$\diamond$: Rotation of the exoplanet is not included.  $\heartsuit$: Non-hydrostatic.\\
\textbf{Acronyms: Boussinesq (BQ), anelastic (AN), equivalent barotropic (EB), primitive equations (PE), Euler equation (EE), Navier-Stokes equation (NS).}\\
\caption{Summary table of atmospheric circulation studies of hot exoplanets using GCMs}
\end{sidewaystable}

\appendix

\section{The Governing Equations of Atmospheric Dynamics}
\label{append:equations}

\subsection{General Form: Navier-Stokes Equation}

The general, governing equation of fluid dynamics is known as the Navier-Stokes equation \citep{vallis06},
\begin{equation}
\frac{D \vec{v}}{D t} = -2\vec{\Omega} \times \vec{v} -\frac{\nabla P}{\rho} + \nu \nabla^2 \vec{v} + \frac{\nu}{3} \nabla\left(\nabla.\vec{v}\right) + \vec{g} + \vec{F}_{\rm drag},
\label{eq:navier}
\end{equation}
where $\vec{v}$ is the velocity vector, $t$ represents the time, $\vec{\Omega}$ is the rotation rate vector, $P$ is the pressure, $\rho$ is the mass density, $\nu$ is the (constant) molecular viscosity, $\vec{g} = -g \hat{z}$ is the surface gravity or acceleration due to gravity, $\hat{z}$ is the unit vector of the vertical spatial coordinate and $\vec{F}_{\rm drag}$ represents the various drag forces per unit mass.  Equation (\ref{eq:navier}) formally expresses the linear conservation of momentum in a fluid.  In atmospheric applications, either the assumption of an inviscid ($\nu=0$) or an incompressible ($\nabla.\vec{v}=0$) fluid is often made.  Enforcing the former and latter assumptions yield the Euler and the incompressible Navier-Stokes equations, respectively.

The Navier-Stokes equation has three scalar components and five variables.  To close the set of equations, we need three equations and one more variable.  One of them is the mass continuity equation,
\begin{equation}
\frac{\partial \rho}{\partial t} + \nabla.\left(\rho\vec{v}\right) = 0,
\label{eq:continuity}
\end{equation}
which formally expresses the conservation of mass.  The thermodynamic equation introduces temperature ($T$) into the system and enforces the conservation of energy.
\begin{equation}
\frac{D T}{D t} = \frac{\kappa T}{P} \frac{D P}{D t} + Q,
\label{eq:thermo}
\end{equation}
where $\kappa$ is the adiabatic coefficient and the term $Q$ represents sources of heating (including from stellar irradiation).  Finally, the assumption of an equation of state, usually that of an ideal gas, ensures an equal number of equations and variables,
\begin{equation}
P = \rho {\cal R} T,
\label{eq:ideal_gas}
\end{equation}
where ${\cal R}$ is the specific gas constant.  The assumption of an ideal gas breaks down in the deep interior of an exoplanet, where pressures are high and degeneracy starts to become important.

\subsection{Boussinesq Approximation}

An approach that is seldom adopted in studies of exoplanetary atmospheres is to apply the Boussinesq approximation, which assumes that density variations are negligible (i.e., incompressibility) except when they are induced by gravity.  

\subsection{Anelastic Approximation}

The anelastic approximation is a slight generalization of the Boussinesq approximation in that we now have the background state of density varying with height.  

\subsection{Shallow Water Approximation}

The shallow-water system of equations governs the dynamics of a single, vertically uniform, hydrostatically balanced layer of constant-density fluid under the assumption that the structures being modeled have wavelengths long compared to the fluid thickness.  By definition, the Boussinesq, anelastic and shallow-water approximations preclude the treatment of acoustic waves (and thus shocks).

\subsection{Equivalent Barotropic Approximation}

The equivalent barotropic approximation \citep{salby90} is similar to the shallow water one as the layer thickness is also a variable of the system, but it is more general in the sense that it allows for a direct calculation of the temperature.  The lower boundary of the system is generalized to an isentropic surface, with constant potential temperature, of arbitrary physical shape.

\subsection{Hydrostatic Primitive Equations}
\label{subsect:primitive}

Originally designed for the study of Earth, there is a reduced form of the set of equations in (\ref{eq:navier}), (\ref{eq:continuity}), (\ref{eq:thermo}) and (\ref{eq:ideal_gas}) that is commonly utilized in the study of atmospheric dynamics and is known as the ``primitive equations".  It involves three sets of approximations.
\begin{itemize}

\item \textbf{Hydrostatic balance:} The assumption that the vertical component of the velocity is much less than the sound speed.  It does \textit{not} mean that the atmosphere is vertically static, i.e., that the vertical velocity is zero.

\item \textbf{Shallow atmosphere:} Let the radial coordinate be represented by $r$.  The lower boundary of the model atmosphere is set to be $r=R$ and the vertical spatial coordinate is $z$.  Thus, we have $r = R + z$.  The ``shallow atmosphere" approximation simply asserts that $r \approx R$ (but retains $dr = dz$).

\item \textbf{Traditional approximation:} The set of governing equations is first written out in spherical coordinates.  It is then assumed that the Coriolis and curvature terms involving the vertical velocity $v_z$ are sub-dominant.  However, the term involving $v_z$ in the $D/Dt$ operator is retained.

\end{itemize}

\end{document}